\documentclass{article}

\usepackage{lineno,hyperref}
\usepackage{amsmath,amssymb}
\usepackage{graphicx}

\usepackage[utf8]{inputenc}
\usepackage[english]{babel}
\usepackage[T2A]{fontenc}

\usepackage{comment} 

\modulolinenumbers[10]
\modulolinenumbers[1]
\newcommand\nnfootnote[1]{%
  \begin{NoHyper}
  \renewcommand\thefootnote{}\footnote{#1}%
  \addtocounter{footnote}{-1}%
  \end{NoHyper}
}

\begin{document}
\title{Dynamics of a 2D lattice of van der Pol oscillators with nonlinear repulsive coupling}
\author{I.A. Shepelev \footnotemark[1], S.S. Muni \footnotemark[2], T.E. Vadivasova\footnotemark[1]}

{\renewcommand\thefootnote{\fnsymbol{footnote}}%
  \footnotetext[1]{Department of Physics, Saratov State University, 83 Astrakhanskaya Street, Saratov, 410012, Russia}\footnotetext[2]{School of Fundamental Sciences, Massey University, Palmerston North, New Zealand}}
\maketitle
\begin{abstract}
We describe spatiotemporal patterns in a network of identical van der Pol oscillators coupled in a two-dimensional geometry.  In this study, we show that the system under study demonstrates a plethora of different spatiotemporal structures including chimera states when the coupling parameters are varied. Spiral wave chimeras are formed in the network when the coupling strength is rather large and the coupling range is short enough. Another type of chimeras is a target wave chimera. It is shown that solitary states play a crucial role in forming an incoherence cluster of this chimera state. They can also spread within the coherence cluster. Furthermore, when the coupling range increases, the target wave chimera evolves to the regime of solitary states which are randomly distributed in space. Growing the coupling strength leads to the attraction of solitary states to a certain spatial region, while the synchronous regime is set in the other part of the system. This spatiotemporal pattern represents a solitary state chimera, which is firstly found in the system of continuous-time oscillators. We offer the explanation of these phenomena and describe the evolution of the regimes in detail.
\end{abstract}

\nnfootnote{Keywords: Spatiotemporal pattern formation, chimera state, van der Pol oscillator, spiral wave, spiral wave chimera, target wave, target wave chimera, solitary state, solitary state chimera, nonlocal coupling}
\nnfootnote{E-mail addresses: I.A. Shepelev (\url{igor\_sar@li.ru}),  S.S. Muni(\url{s.muni@massey.ac.nz}), T.E. Vadivasova (\url{vadivasovate@yandex.ru})}
\date{}

\section*{\label{sec:intro}Introduction}

The study of the self-organization phenomena in complex multicomponent systems in the form of oscillatory networks and ensembles is one of the most relevant directions in the nonlinear dynamics and related  disciplines  \cite{Kuramoto-1984, Nekorkin-2002, Osipov-2007, Barrat-2008, Boccalett-2018}. The multicomponent systems with different dynamics of individual elements are models of many real systems both in nature and technology. Their examples represent neuronal network, population of living organisms, transport and computer networks and so on. The features of the element to element interaction significantly impacts the dynamics of the multicomponent systems (networks and ensembles), the synchronization effects, the formation of different types of waves and cluster structures. The coupling topology plays an important role \cite{Genio-2016, Bera-2017, Yamamoto-2018, Belych-2018}. However, the features of the element to element coupling, namely a type of the coupling function, is very important too. Generally, the linear inertialess coupling is considered in most of the models of ensembles and complex networks. It increases the dissipation and, usually, is called dissipative coupling. In the case of identical systems, the dissipative coupling leads to a regime of the complete synchronization, when oscillations of the interacting systems are in-phase. At the same time, a type of the coupling in real complex systems can be different. It can be dissipative and inertial \cite{Kuznetsov-2009}, nonlinear \cite{Yang-2016, Petereit-2017}, memristive \cite{Volos-2015, Xu-2018} and with delayed feedback \cite{Yeung-1999, Liu-2007, Schmidt-2012, Kaue-2012}. A special type of coupling represents the repulsive interaction, when the coupling coefficient is negative. In general, the repulsive coupling impedes the emergence of in-phase oscillations. An interest to the study of features of systems with repulsive interaction is explained by the fact that this type of coupling takes place in issues of biophysics and neurodynamics \cite{Balazsi-2001, Wang-2001, Yanagita-2005, Rabinovich-2006, Ullner-2007}. There are a lot of works, where different types of repulsive and mixed coupling have been studied. Systems of two or three self-oscillators with the periodic or chaotic dynamics have been investigated for the cases of repulsive and mixed coupling in \cite{Chen-2009, Hens-2014, Astakhov-2016, Zhao-2018, Dixit-2019}. It has been shown that the repulsive coupling in self-oscillating systems usually leads to the amplitude or oscillatory death, while this coupling can induce oscillations in excitable systems \cite{Yanagita-2005}. There are a number of works, in which ensembles of oscillators have been studied for the cases of both local \cite{Balazsi-2001, Ullner-2007} and global \cite{Tsimring-2005} repulsive coupling, and for the case of mixed attractive and repulsive coupling \cite{Hong-2011, Hens-2013, Nandan-2014, Maistrenko-2014}. Moreover, systems with the local and global repulsive coupling with the delayed feedback has been considered in \cite{Wang-2001, Bera-2016}. The repulsive coupling in the ensembles of self-oscillators leads to the amplitude and oscillatory death \cite{Ullner-2007, Hens-2013, Nandan-2014}. An ensemble of phase oscillators demonstrates different synchronous regimes \cite{Tsimring-2005}, desynchronization, partial synchronization, traveling waves \cite{Hong-2011}. Oscillatory ensembles with mixed  attractive (dissipative) and repulsive coupling show regimes with the complex spatiotemporal dynamics, namely the solitary states \cite{Maistrenko-2014} and chimera-like structures \cite{Mishra-2015}.

Despite these works, there are still many issues concerning the dynamics of systems with repulsive coupling. The dynamics of self-oscillator ensembles is yet to be explored for the case of repulsive nonlocal interaction between the oscillators. It is known that the nonlocal coupling with a limited number of coupled neighbors in ensembles of identical oscillators of different types, from the phase oscillators to stochastic excitable systems, can lead to the formation of chimera states for certain values of parameters. These states presents the cluster structures, including patterns with the coherent and incoherent behavior \cite{Kuramoto-2002, Abrams-2004, Omelchenko-2011, Hagerstrom-2012, Martens-2013, Zakharova-2014, Panaggio-2015, Rybalova-2019, Shepelev-2019, Zakharova-2020}. The question arises about the presence of chimera states and other complex structures in ensembles with the nonlocal repulsive coupling. A wide variety of spatiotemporal dynamics can be excepted in a 2D lattice of the self-oscillators with the both local and nonlocal repulsive coupling. Another unexplored issue is the influence of the nonlinear character of repulsive coupling, which can be represented by a two-pole with negative conductivity. This coupling element should have the current-voltage curve with a falling section and be an active element which adds the negative dissipation to the system. Furthermore, this element is essentially nonlinear. Hence, introducing the constant coupling coefficient is an approximate description of the coupling function, which is valid only when the amplitude is low. Generally, it is necessary to take into account the influence of the nonlinearity of repulsive coupling on the oscillator dynamics.

In the present work, we simulate the dynamics of a 2D lattice of the identical van der Pol (vdP) oscillators, which are coupled by the nonlocal nonlinear repulsive coupling. The aim of this study is to establish the features of the formation of spatial structures in the lattice with the nonlocal nonlinear repulsive interaction when different coupling parameters are varied, namely the coupling coefficient and the coupling range. We also show the role played by the coupling nonlinearity in the formation of different spatiotemporal dynamics of the lattice.

\section{\label{sec:system}System under study}

\begin{figure}[!ht]
\centering
  \includegraphics[width=\linewidth]{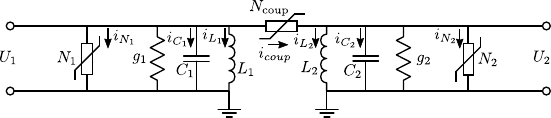}
\caption{Circuit diagram of two electronic oscillators coupled through a tunnel diode.}
\label{fig:circuit_diag}
\end{figure}

At first we consider an electronic circuit of two self-oscillators coupled through an active nonlinear element 
with the negative conductivity  (for example  a tunnel diode). An equivalent circuit design is illustrated in fig.\ref{fig:circuit_diag}.
%
%
\\
Here $g_{1,2}$, $L_{1,2}$, $C_{1,2} $ and $N_{1,2}$ are linear conductivity, inductors, capacitors and nonlinear elements of the two self-oscillators, respectively. $N_{coup}$ is an element of the nonlinear coupling. The currents $i_{N_{1,2}} $ and $i_{coup} $ through the nonlinear elements $N_{1,2}$ and $N_{coup}$ are described by the following expression:
\begin{equation}
\begin{array}{l}
i_{N_{1,2}} = -\alpha_{1,2}U_{1,2} + \beta_{1,2}U_{1,2}^{3},\\[8pt]
i_{coup} = -\gamma U + \delta U^3,
\end{array}
\label{eq1}
\end{equation}
where $\alpha_{1,2}$, $\beta_{1,2}$, $\gamma$, $\delta$ are certain positive parameters.
The chosen form of the current-voltage characteristics of the nonlinear elements $N_{1,2}$ corresponds to van der Pol oscillators. 
The Kirchhoff equations of this circuit take the following form:

\begin{equation}
\begin{array}{l}
    C_{1} \dfrac{dU_{1}}{dt_1} + i_{L_1} + g_{1}U_{1} + i_{N_{1}} + i_{coup} = 0\\[10pt]
    C_{2} \dfrac{dU_{2}}{dt_1} + i_{L_2} + g_{2}U_{2} + i_{N_{2}} - i_{coup} = 0,\\
    \dfrac{di_{1,2}}{dt_1} = \dfrac{1}{L_{1,2}} U_{1,2},~~~~~
    C_{1,2} \dfrac{dU_{1,2}}{dt_1} = i_{C_{1,2}}. \\
\end{array}
\label{eq2}
\end{equation} 
Here $t_1$ is real time. The following equations for the voltages $U_{1,2}$ can be obtained when the currents $i_{L_{1,2}}$, $i_{C_{1,2}}$, $i_{N_{1,2}} $ and $i_{coup} $ are excluded:
\begin{equation}
\begin{array}{l}
    \dfrac{d^{2}U_{1}}{dt_1^{2}} - \left( \dfrac{\alpha_{1} - g_{1}}{C_{1}} - \dfrac{3 \beta_{1}}{C_{1}} U_{1}^{2}\right) \dfrac{dU_{1}}{dt_1} + \omega_{01}^{2} U_{1} = \\ = - \left( \dfrac{\gamma}{C_{1}} - \dfrac{3 \delta}{C_{1}} \left( U_{1} - U_{2}\right)^{2}\right) \left( \dfrac{dU_{2}}{dt_1} - \dfrac{dU_{1}}{dt_1}\right)\\[15pt]
    \dfrac{d^{2}U_{2}}{dt_1^{2}} - \left( \dfrac{\alpha - g_{2}}{C_{2}} - \dfrac{3 \beta_{2}}{C_{2}} U_{2}^{2}\right) \dfrac{dU_{2}}{dt_1} + \omega_{02}^{2} U_{2} = \\ = -\left( \dfrac{\gamma}{C_{2}} - \dfrac{3\delta}{C_{2}} \left( U_{1} - U_{2}\right)^{2}\right) \left( \dfrac{dU_{1}}{dt_1} - \dfrac{dU_{2}}{dt_1}\right),
\end{array}
\label{eq3}
\end{equation}
where $\omega_{01}=(L_1 C_1)^{-1/2}$ and $\omega_{02} = (L_2 C_2)^{-1/2}$ are the parameters specifying frequencies of the two self-oscillators. Let $C_{1} = C_{2}  =C, \alpha_{1} = \alpha_{2} = \alpha, \beta_{1} = \beta_{2} = \beta$ and $L_{1} \neq L_{2}$.
Using the substitution $t = \omega_{0}t_{1}$, $\sqrt{\frac{3\beta}{c\omega_{0}}} U_{1,2} = x_{1,2}$, where $\omega_0$ is some fixed frequency,
we obtain the following system of equations in dimensionless variables:
\begin{equation}
\begin{array}{l}
    \ddot{x}_{1} - \left( \varepsilon - x_{1}^{2} \right)\dot{x}_{1} + \omega_{1}^{2} x_{1} = - \left( k - m(x_{1}  -x_{2})^{2} \right) \left( \dot{x}_{2} - \dot{x}_{1}\right)\\[10pt]
    \ddot{x}_{2} - \left(\varepsilon - x_{2}^{2} \right) \dot{x}_{2} + \omega_{2}^{2} x_{2} = -\left(k - m(x_{1} - x_{2})^{2} \right) \left( \dot{x}_{1} - \dot{x}_{2}\right)
\end{array}
\label{eq4}
\end{equation}
where $\varepsilon = \frac{\alpha - g}{C \omega_0}$,  $\omega_{1} = \frac{\omega_{01}}{\omega_{0}}$, $\omega_{2} = \frac{\omega_{02}}{\omega_{0}}$, $k = \frac{\gamma}{\omega_{0}C}$, $m = \dfrac{\delta}{\beta}$. 
We rewrite the model under study in the form of a system of differential equations of the first order:
\begin{equation}
\begin{array}{l}
    \dot{x}_{1} = y_{1},\\
    \dot{y}_{1} = \left(\varepsilon - x_{1}^{2} \right) y_{1} - \omega_{1}^{2} x_{1} - \left(k - m(x_{1} - x_{2})^{2} \right) \left( y_{2} - y_{1}\right),\\
    \dot{x}_{2} = y_{2},\\
    \dot{y}_{2} = \left( \varepsilon - x_{2}^{2}\right)y_{2} - \omega_{2}^{2} x_{2} - \left( k - m(x_{1} - x_{2})^{2}\right) \left( y_{1} - y_{2}.\right)
\end{array}
\label{eq5}
\end{equation}
If $\omega_1 = \omega_2$ then both the oscillators are completely identical. 

Similarly, we can write the equations for 2D lattice of $N $ identical van der Pol oscillators with the nonlocal repulsive nonlinear coupling, which we are going to study:
%
\begin{equation}
\begin{array}{l}
\dot{x}_{i,j}(t) = y_{i,j},
\\
\dot{y}_{i,j}(t) =  (\varepsilon- x_{i,j}^2) y_{i,j} - \omega^2 x_{i,j} - 
\sum\limits_{\tiny \begin{aligned}& k=i-P \\[-4pt] & p=j-P\end{aligned}}^{\tiny \begin{aligned}& i+P \\[-4pt] & j+P\end{aligned}}
\left(k-m\left(x_{k,p} - x_{i,j}\right)^2\right)\left(y_{k,p} - y_{i,j}\right),\\[12pt]
i,j=1,...N,
\end{array}
\label{eq:grid_1}
\end{equation}
%
The double index of dynamical variables $ x_{i, j} $ and $ y_{i, j} $ with $ i,j = 1, ..., N $ determines the position of an element in the two-dimensional lattice. All the oscillators are identical with respect to parameters and each of them is coupled with all the lattice elements from a square with side $ (1 + 2P) $ in the center of which this element is located. The integer $ P $ defines the nonlocal character of coupling and is called the coupling range. The case of $P=1 $ corresponds to the local coupling, while $P=N/2 $ is the case of  global coupling, when each element interacts with the whole system. It determines the  number of neighbors $ Q = (1 + 2P) ^ 2-1 $, which each element is coupled with.  
Parameter $k $ is the coefficient of the linear repulsive coupling term while $m $ is the coefficient of the attractive nonlinear coupling term. 

Two components of coupling can be distinguished from the system (\ref{eq:grid_1}): linear repulsive and nonlinear attractive. With this in mind, the equations under study take the form
%
\begin{equation}
\begin{array}{l}
\dot{x}_{i,j}(t) = y_{i,j}  ,
\\
\dot{y}_{i,j}(t) =  (\varepsilon- x_{i,j}^2) y_{i,j} - \omega^2x_{i,j} - 
\dfrac{\sigma}{Q}
\sum\limits_{\tiny \begin{aligned}& k=i-P \\[-4pt] & p=j-P\end{aligned}}^{\tiny \begin{aligned}& i+P \\[-4pt] & j+P\end{aligned}}
\left(y_{k,p} - y_{i,j}\right) +\\
 ~~~~~~~~~~+m \sum\limits_{\tiny \begin{aligned}& k=i-P \\[-4pt] & p=j-P\end{aligned}}^{\tiny \begin{aligned}& i+P \\[-4pt] & j+P\end{aligned}}
\left(x_{k,p} - x_{i,j}\right)^2\left(y_{k,p} - y_{i,j}\right),\\[12pt]
i,j=1,...N.
\end{array}
\label{eq:grid}
\end{equation}
Here the linear repulsive coupling coefficient $\sigma$ is introduced, which is reduced to the number of coupled neighbors $Q $,  while the nonlinear attractive term does not depend on the parameters $P $ and $\sigma $. For simplicity, we will call the parameter $\sigma $ as the coupling strength. It can be seen from the eq.\eqref{eq:grid} that the ratio between the linear and nonlinear parts has noticeably changed when the coupling range $P $ is varied. When the value of $P$ is low, the linear part prevails over the nonlinear one. However the contribution of the nonlinear part increases intensively with the elongation of the coupling range. 
Let us consider the following example. Let a value of the coupling strength is chosen as $\sigma=0.8 $. The value of $k $ in the eq.\eqref{eq:grid_1} is $k=\sigma/((2P+1)^2-1) $. Thus, $k=0.1 $ ($\frac k m=5 $) when $P=1 $ and already $k=0.01 $ ($\frac{k}{m}=0.5 $) when $P=4 $. For this reason, we expect a significant change in the system dynamics with an elongation of the coupling range due to the increasing influence of the coupling nonlinearity.
 
We will assume that the boundary conditions for (\ref{eq:grid}) are toroidal, i.e. periodic in both directions:
%
\begin{equation}
\begin{cases}
x_{0,j}(t) = x_{N,j}(t),\\
x_{i,0}(t) = x_{i,N}(t),\\
\end{cases}
\begin{cases}
y_{0,j}(t) = y_{N,j}(t),\\
y_{i,0}(t) = y_{i,N}(t),
\end{cases}
\label{eq:boundaries}
\end{equation}
%
Initial conditions for all considered cases are the random values of the variables with an uniform distribution within $x_0 \in [-2,2],~y_0 \in [-2,2] $. The system equations are integrated using the Runge-Kutta 4th order method with a time step of  $dt_1=0.001$. All the regimes under study are obtained after the transient process of $t_{\rm trans}=6000 $ time units.

\section{Basic dynamical regimes of the lattice}

Now we start to explore the dynamics of the lattice \eqref{eq:grid} when the parameters $\sigma$ and $P $ are varied and the control parameters are fixed as $m=0.02 $, $\varepsilon=2.1 $, and $\omega=2.5 $. Note that a value of $\omega $ does not play a sufficient role and can be excluded by the variable replacement. 
Thus, the nonlinear part of the coupling is significantly lower than the linear part. At the same time, we show that even a small nonlinear addition into the coupling leads to a qualitative change in the system under study.
 
\begin{figure}[!ht]
\centering
  \includegraphics[width=0.6\linewidth]{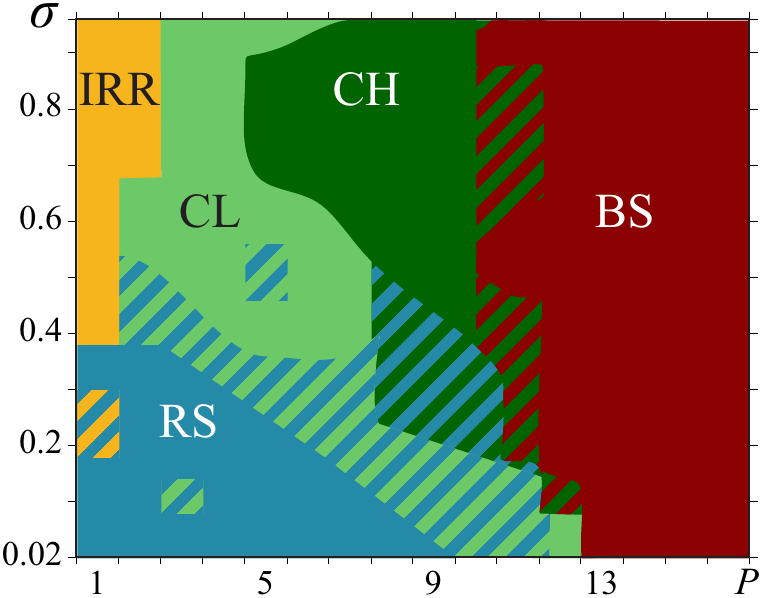}
\caption{(Color online) Regime diagram for the lattice \eqref{eq:grid} in the ($P,~\sigma$) parameter plane when $m=0.02 $, $\varepsilon=2.1 $ and $\omega=2.5$.
Region IRR corresponds to the complete incoherence; region RS relates to the regime of regular states; CL is the region of existence of chimera-like structures; CH corresponds to the chimera states; TS is the region of  two-state structures. The regions of coexistence of different regimes are shown by alternating strips of corresponding colors (tones).}
\label{fig:regime_map}
\end{figure}

We plot a regime diagram for the lattice \eqref{eq:grid} in the ($r,~\sigma$) parameter plane within the range $\sigma \in [0,1]$ and $P \in [1,16] $ as shown in Fig.~\ref{fig:regime_map}(a). 
A sequence of randomly distributed initial conditions within the intervals $x_{i,j} \in [-2,2]$ and $ y_{i,j} \in [-2,2]$ are used to construct the regime diagram. The dashed regions correspond to the case of multistability in the lattice, when two different steady regimes are observed for various sets of the initial conditions.

An important feature of the system  \eqref{eq:grid} with the repulsive coupling is the absence of any type of propagating wave regimes. There are no traveling waves, spiral and target waves for any values of $\sigma $, $P $, and initial conditions.
 Only standing waves are realized in the lattice \eqref{eq:grid}.
Spatially homogeneous regimes with the complete synchronization of oscillators are also not typical for the system \eqref{eq:grid}.The repulsive coupling increases the total energy. This leads to dissynchronization between the system elements, what is most noticeable when the coupling range is very short. This effect is observed in the region $IRR  $ in the regime diagram for the cases $P=1 $ (local coupling) and $P=2 $. The spatial structure becomes more and more complex with increase in the coupling strength. Elongation of the coupling range significantly attenuates this effect and the appearance of the incoherent structures is observed only for large values of $\sigma>0.7 $. When the coupling strength is sufficiently low  and coupling range is short ($P<13 $) regular spatiotemporal states are observed in the lattice. These states are characterized by a piecewise smooth spatial profile and the very similar periodic oscillations of all the lattice elements.  They exist within the region $RS$ in the regime diagram in fig.\ref{fig:regime_map}. Unlike the case of local coupling, growth of $\sigma$ leads to the formation of the incoherence only within a certain spatial region while the structure in the rest part of a lattice remains regular. Thus, the state with coexisting coherence and incoherence domains appear in the system what are typically the chimera states. However, we divide these states into two groups, namely the regime of chimera-like states ($CL $ region) and  the regime of chimeras ($CH $ region). A $(P,\sigma)$ tuple is classified as $CL$ state if the number of incoherent elements constitute less than $10\% $ of the lattice and do not form the incoherence cluster. 
The regime diagram illustrates that the regime of chimera-like structures (region $CL $) leads to the evolution of regular structures with increase in the coupling strength. In turn, the region of chimera states is observed for the longer coupling range than $CL$ region  and expands with an increase in the values of $\sigma $. Further elongation of $P $ leads to a switch of the system to "two-state" regime (region $TS $), when all the oscillators are irregularly distributed between two characterized states. Only this regime is realized for the long coupling range ($P \geqslant 13 $). Next we explore all the main regimes in detail.

\subsection{Regular spatiotemporal structures}

\begin{figure}[!ht]
\centering
\hspace{-1mm}\parbox[c]{.345\linewidth}{ 
  \includegraphics[width=\linewidth]{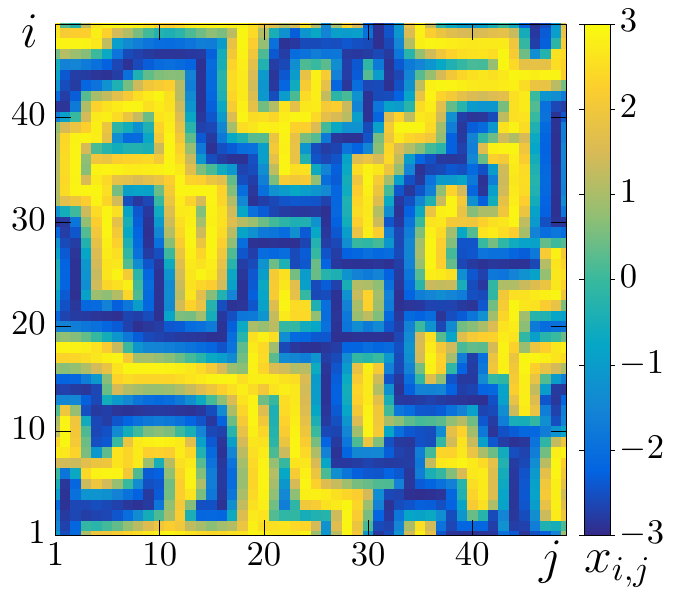}
    \vspace{-9.5mm} \center (a)
      \includegraphics[width=\linewidth]{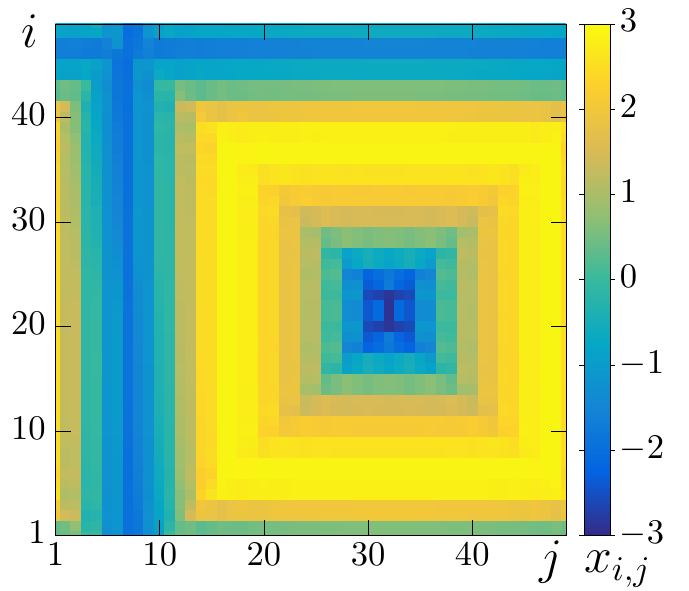}
    \vspace{-9.5mm} \center (d)
}
\hspace{-2mm}\parbox[c]{.345\linewidth}{
\includegraphics[width=\linewidth]{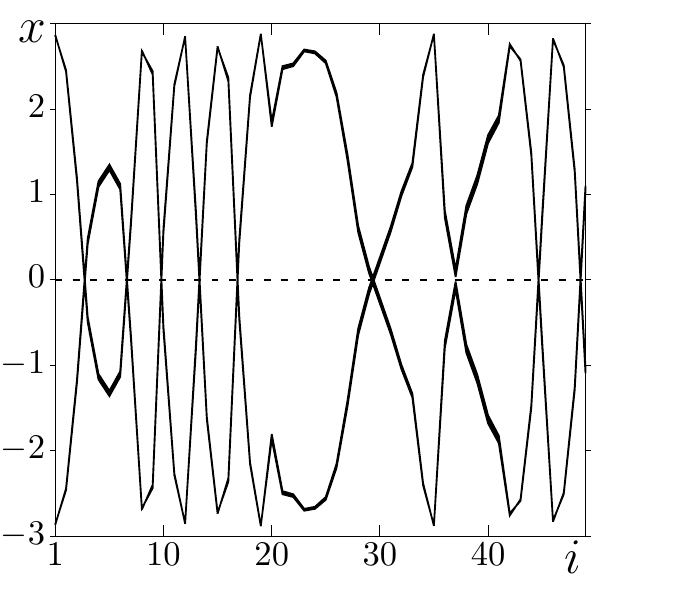}
\vspace{-9.5mm}\center (b)
  \includegraphics[width=\linewidth]{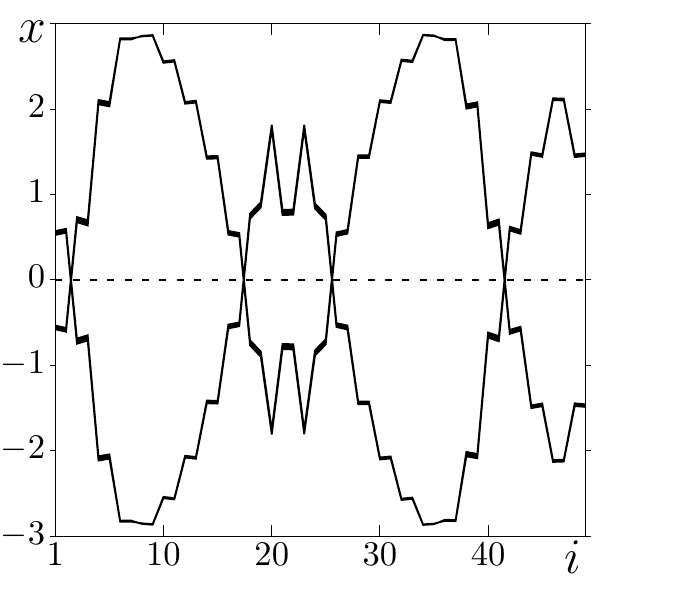}
    \vspace{-9.5mm} \center (e)
}
\hspace{-5mm}\parbox[c]{.345\linewidth}{
\includegraphics[width=\linewidth]{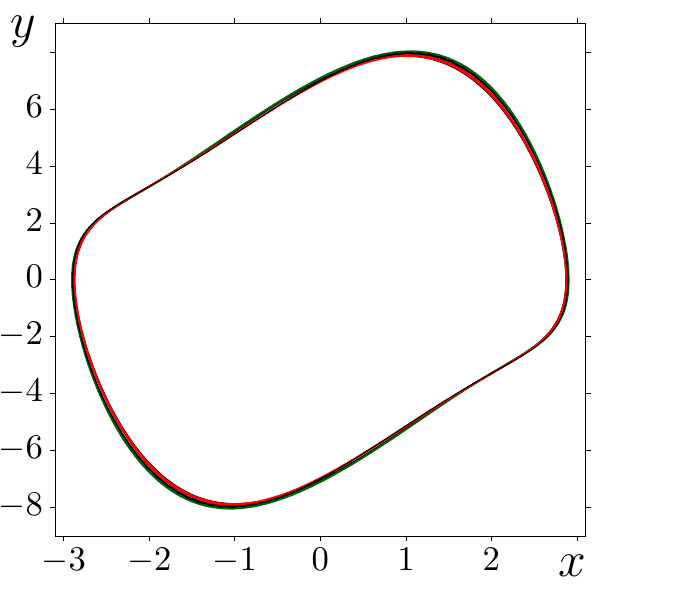}
\vspace{-9.5mm}\center (c)
  \includegraphics[width=\linewidth]{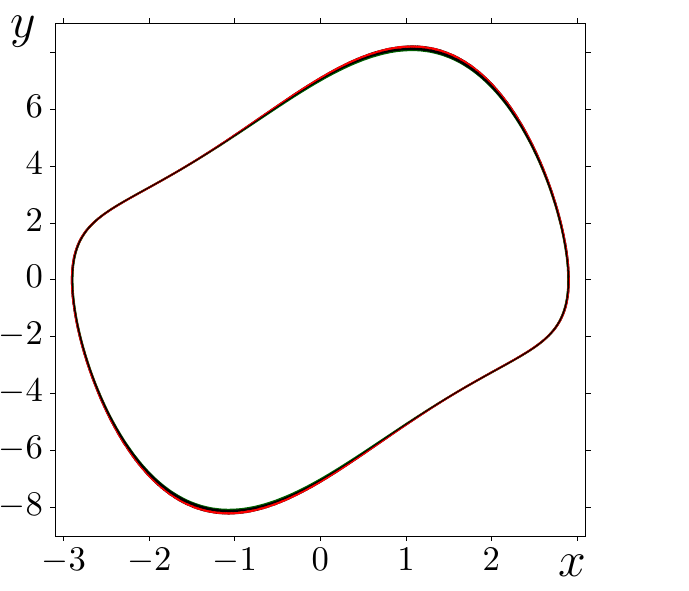}
    \vspace{-9.5mm} \center (f)
}
\caption{(Color online) Regular spatiotemporal states in the system \eqref{eq:grid} (region $RS $ in the regime diagram in fig.\ref{fig:regime_map}). The case of local coupling $P=1 $ with $\sigma=0.2 $ is illustrated by panels  (a)-(c), and the case of $P=2 $ and $\sigma=0.18 $ corresponds to panels (d)-(f). (a), (d) snapshots of the system states; sets of 20 instantaneous spatial cross-sections shown every the half period for $j=10 $ (b) and $j=31 $ (e); projections of the phase trajectories for the $i=7, j=20 $th (solid black line), $i=30, j=22 $th (solid green line), $i=37, j=15 $th (solid red line), $i=44, j=35 $th (dotted black line) elements (c) and for the $i=17, j=31 $th (solid black line), $i=25, j=31 $th (solid green line), $i=26, j=31 $th (solid red line), $i=41, j=20 $th (dotted black line) elements (f). Parameters: $m=0.02 $, $\varepsilon=2.1 $, $\omega=2.5$, $N=50$.} 
\label{fig:REG}
\end{figure}

We study the regime of regular spatiotemporal patterns realized in the region $RS$ of the regime diagram in fig.\ref{fig:regime_map}. Spatial profiles of the structures for the cases of local and nonlocal coupling are significantly different. An example of the structure for $P=1 $ is illustrated by a snapshot of the system state in fig.\ref{fig:REG}(a). The shape of this structure is sufficiently complex, however the spatial profile is always smooth and the instantaneous states of the adjacent oscillators are very similar.  Its spatiotemporal dynamics is depicted in fig.\ref{fig:REG}(b) by a set of instantaneous cross sections of the two-dimensional spatial profile shown after every the half period $T/2 $. It can be seen that the spatial profile repeats itself after each period $T=2.83 $ time units. This indicates that a motion of the wavefront is absent and the regime represents the standing wave, and also that oscillations of all the elements are periodic.
%
The regular character of oscillations is also shown by projections of phase trajectories for different elements of the lattice in fig.\ref{fig:REG}(c). Furthermore, all the oscillators are characterized by almost the same attractors. It should be noted that here and further we do not use the terms "an attractor" and "a limit cycle" in a sense of an attractor in the phase space of the multidimensional dynamical system \eqref{eq:grid}, but in a sense of an attracting manifold in a plane of the variables of an individual oscillator. Hence, there is only one steady state in the system.

When the coupling becomes nonlocal ($P=2 $), the shape of the spatial profile undergoes a significant change. A typical spatial structure in the lattice \eqref{eq:grid} for the nonlocal coupling is depicted by a snapshot of the system state in fig.\ref{fig:REG}(d). The distinctive feature of this state is its square step-structure. This structure represents a step pyramid with a deep recess in the center. Apparently, the square shape of the steps is due to the coupling geometry, namely each oscillator is coupled with all the oscillators from a square with the edge $(2P+1) $. A set of  spatial cross-section intersecting the pyramid center is shown in fig.\ref{fig:REG}(e). The spatial cross sections are illustrated after every the half period $T$, which is equal to $T=2.83 $ time units. It is seen that the spatial structure regularly "breathes" in time. Fig.\ref{fig:REG}(f) demonstrates projections of the phase trajectories for different oscillators of the lattice. The plot indicates that all the oscillators are characterized by the same projection in a form of the closed curve. Thus, the oscillators located on different steps are distinguished from each other by only instantaneous phase, while the oscillators on the same step have the very similar instantaneous phase and amplitude. It should be noted, that this type of spatiotemporal structure is unique for a lattice of the coupled van der Pol oscillators and is never observed for the case of attractive coupling. Apparently, this is a result of the repulsive interaction between elements. 

\subsection{Irregular structures}

\begin{figure}[!ht]
\centering
\parbox[c]{.4\linewidth}{
\vspace{-9.5mm}\center $P=1 $ 
  \includegraphics[width=\linewidth]{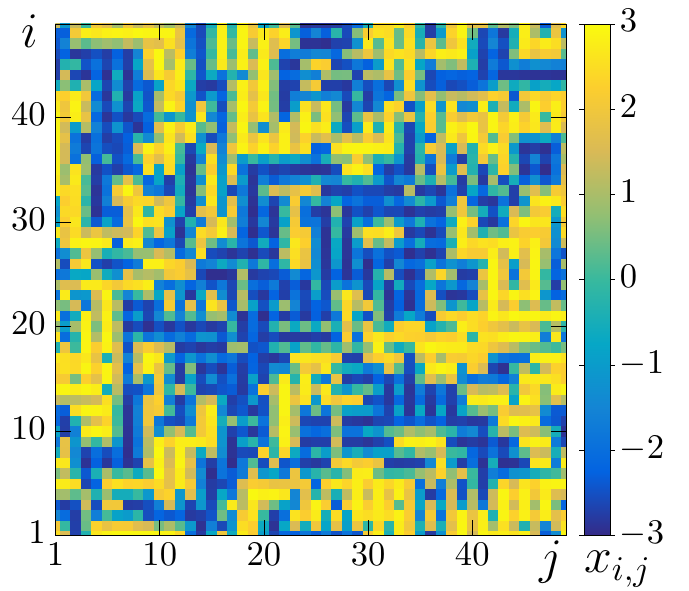}
    \vspace{-9.5mm} \center (a)
\includegraphics[width=\linewidth]{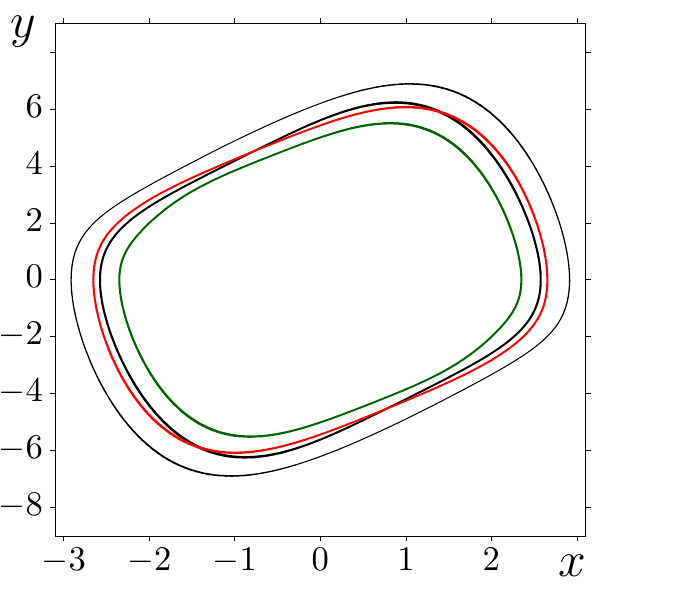}
\vspace{-9.5mm}\center (c)
}
\parbox[c]{.4\linewidth}{
\vspace{-9.5mm}\center $P=2 $ 
  \includegraphics[width=\linewidth]{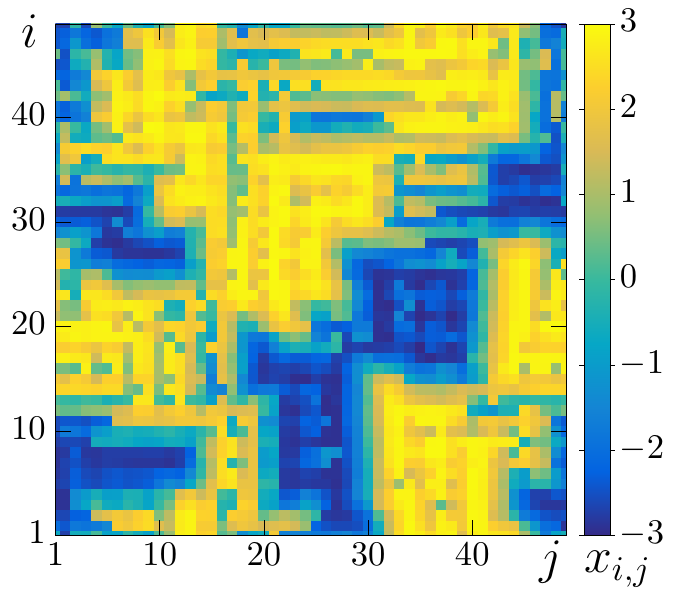}
   \vspace{-9.5mm} \center (b)
\includegraphics[width=\linewidth]{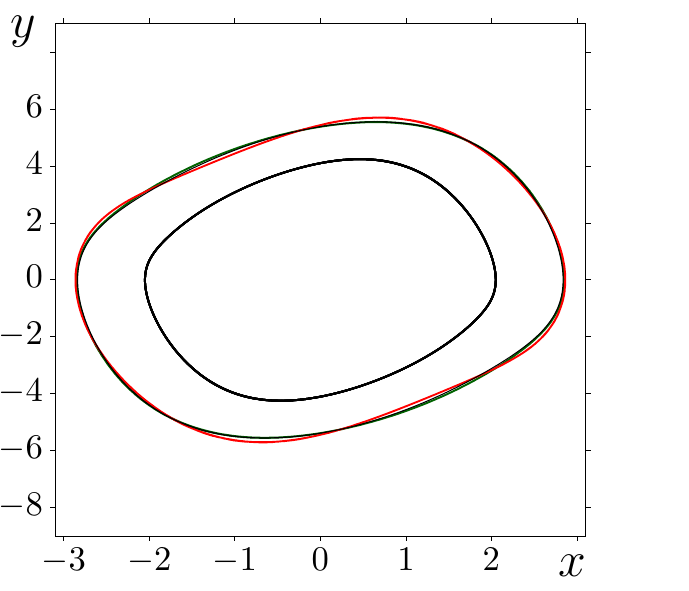}
 \vspace{-9.5mm} \center (d)
}
\caption{(Color online) Irregular structures in the system \eqref{eq:grid} (region $IRR $ in the regime diagram in fig.\ref{fig:regime_map}). The case of local coupling $P=1 $ and $\sigma=0.44 $ is illustrated by panels  (a), (c), and the case of $P=2 $ and $\sigma=0.84 $ corresponds to panels (b), (d). (a) and (b) snapshots of the system states; projections of the phase trajectories for (c) the $i=7, j=20 $th (solid black line), $i=9, j=35 $th (solid green line), $i=16, j=21 $th (solid red line), $i=18, j=21 $th (dotted black line) elements  and (d) shows projections for the $i=4, j=31 $th (solid black line), $i=11, j=17 $th (solid green line), $i=16, j=11 $th (solid red line), $i=31, j=21 $th (dotted black line) elements. Parameters: $m=0.02 $, $\varepsilon=2.1 $, $\omega=2.5$, $N=50 $.}
\label{fig:IRR}
\end{figure}

An increase in the coupling strength leads to the complication of spatial structures. This effect is most pronounced in the case of a very short coupling range $P =1$ and $P=2 $. Moreover, for the case of $P=2 $ this regime is observed for significantly higher values of $\sigma $ than when the coupling is local and is not realized for the longer coupling range within $\sigma \in [0,1] $. Examples of these states are shown in fig.\ref{fig:IRR}(a) for $P=1 $ and in fig.\ref{fig:IRR}(b) for $P=2 $. It can be seen that the shape of spatial profiles becomes significantly complex. At the same time the rectangular geometry of patterns is preserved. Further growth of the coupling strength leads to an increase in the irregularity of the spatiotemporal structures. 
Our study of this regime shows that a growth in the strength of repulsive coupling leads to qualitative and quantitative changes in the system dynamics, namely the lattice \eqref{eq:grid} becomes highly multistate. If for the previous case the all the lattice elements are characterized by the same closed curve in the common $x-y $ plane (which we conditionally call as a \textit{limit cycle}) then for the irregular structures oscillations of different elements are already related to various limit cycles. This effect is clearly seen in the projections of the phase trajectories in fig.\ref{fig:IRR}(c),(d). When the coupling is local, each chosen element is characterized by unique limit cycle. For the case of $P=2 $, this effect is attenuated but it is possible to detect at least three various limit cycles. Hence, growth of the strength of repulsive coupling $\sigma $ leads to the emergence of new stable solutions and the system becomes multistable. Apparently, an increase in the contribution of nonlinear attractive coupling with growth of $P $  reduces a number of the states of lattice elements.

\subsection{Chimera-like structures}

\begin{figure}[!ht]
\centering
\hspace{-1mm}\parbox[c]{.345\linewidth}{ 
  \includegraphics[width=\linewidth]{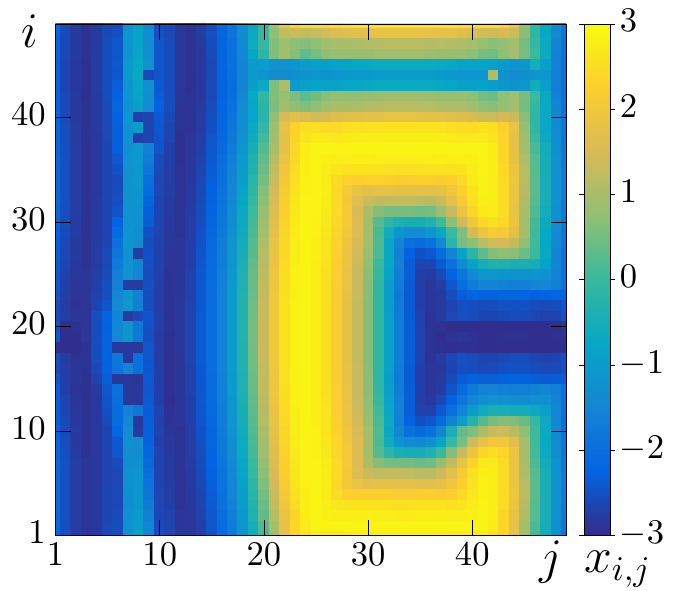}
    \vspace{-9.5mm} \center (a)
      \includegraphics[width=\linewidth]{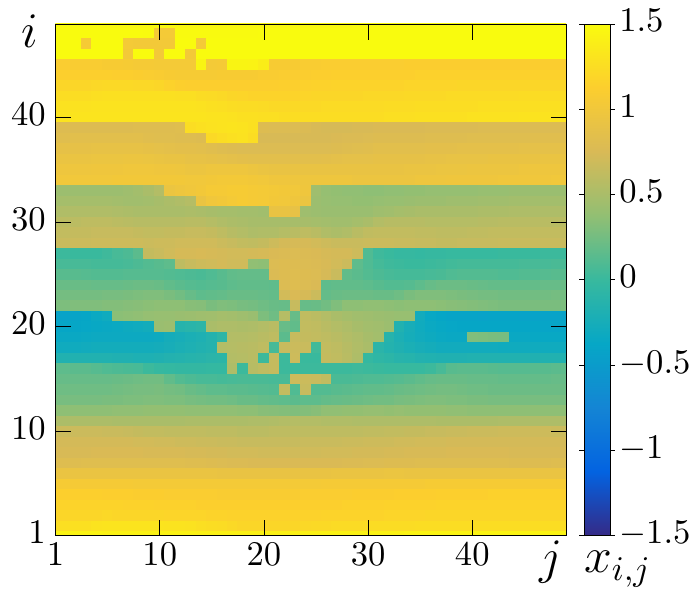}
    \vspace{-9.5mm} \center (d)
}
\hspace{-1mm}\parbox[c]{.345\linewidth}{
\includegraphics[width=\linewidth]{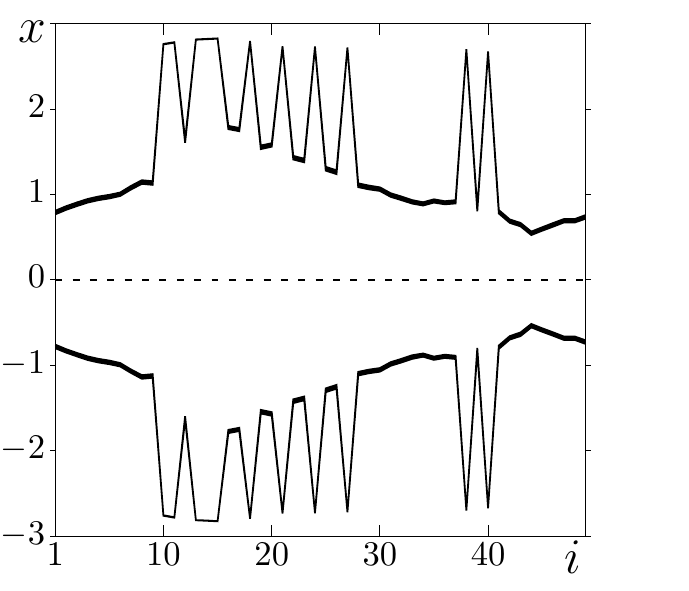}
\vspace{-9.5mm}\center (b)
  \includegraphics[width=\linewidth]{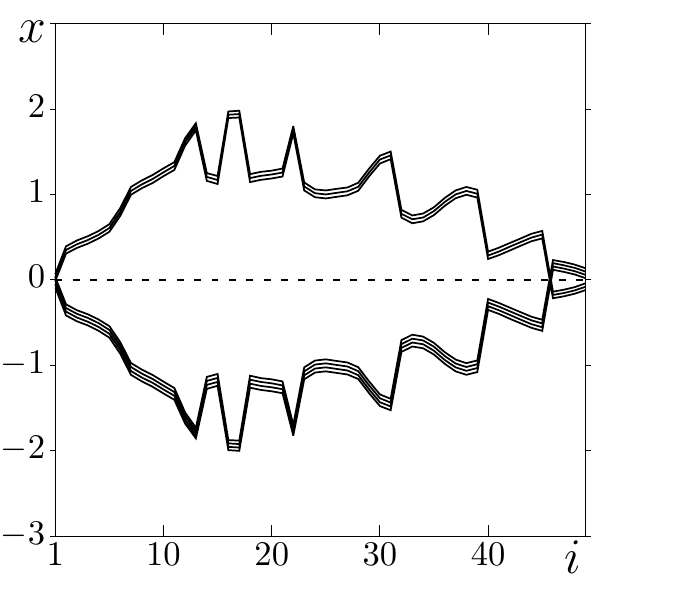}
    \vspace{-9.5mm} \center (e)
}
\hspace{-5mm}\parbox[c]{.345\linewidth}{
\includegraphics[width=\linewidth]{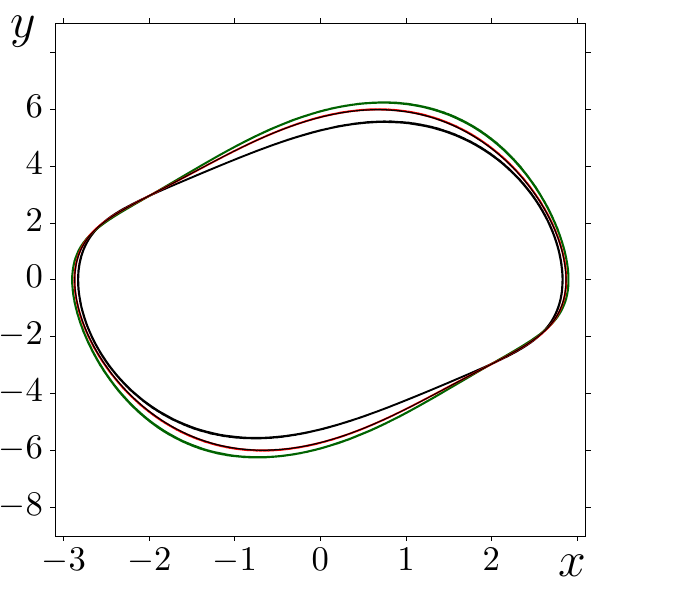}
\vspace{-9.5mm}\center (c)
  \includegraphics[width=\linewidth]{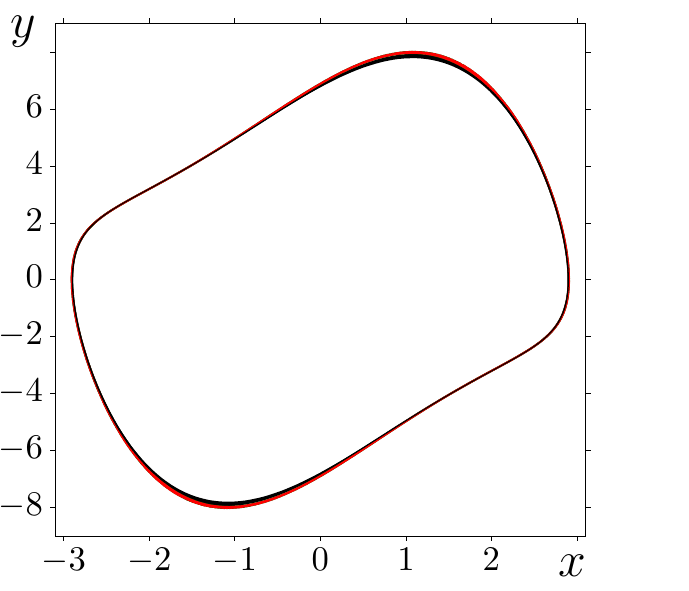}
    \vspace{-9.5mm} \center (f)
}
\caption{(Color online) Chimera-like states in the system \eqref{eq:grid} (region $CL $ in the regime diagram in fig.\ref{fig:regime_map}). The case of $P=3 $ and $\sigma=0.82 $ is illustrated by panels  (a)-(c), and the case of $P=6 $ and $\sigma=0.24 $ corresponds to panels (d)-(f). (a), (d) snapshots of the system states; sets of 20 instantaneous spatial cut-offs shown every the half period for (b) $j=9 $  and (e) $j=25 $; projections of the phase trajectories for (c) the $i=15, j=9 $th (solid black line), $i=20, j=9 $th (solid green line), $i=21, j=21 $th (solid red line), $i=27, j=27 $th (dotted black line) elements  and for (f) the $i=14, j=14 $th (solid black line), $i=17, j=24 $th (solid green line), $i=20, j=24 $th (solid red line), $i=40, j=40 $th (dotted black line) elements. Parameters: $m=0.02 $, $\varepsilon=2.1 $, $\omega=2.5$, $N=50 $.} 
\label{fig:CL}
\end{figure}

Now we discover the spatiotemporal behavior of the regime realized in region $CL $ of the regime diagram in fig.\ref{fig:regime_map}. For this regime contribution of the nonlinear attractive coupling noticeably increases. The spatiotemporal structures represents mixing of two previously described structures, namely the most part of the lattice in partly coherent state, while certain groups of elements oscillate asynchronously with adjacent neighbors. Examples of these structures are illustrated in figs.\ref{fig:CL}(a) and (d) for two different values of $P $. It is visible that the consequence of elongation of the coupling range is the disappearance of the square step-structure. There are a small number of elements which oscillate incoherently (the small incoherence clusters) and  the rest part of system oscillators, which form regular structures (the coherence cluster). The temporal dynamics is regular and is presented by a sets of spatial cross-section through the incoherence clusters in figs.\ref{fig:CL}(b) and (e) for $P=3 $ and $P=6 $, accordingly. The system states are shown every half period $T/2 $. The instantaneous states of oscillators in the incoherence domain are noticeably different from the states of the oscillators from the coherence cluster. Oscillations of all the oscillators are periodic similar to the previous cases. 
As it has been shown above, an increase in the coupling strength $\sigma $ leads to the emergence of new states of individual oscillators of the system under study. At the same time, the elongation of the coupling range $P $ reinforces a contribution of the nonlinear attractive coupling term. Apparently, the nonlinearity of coupling decreases the effect of birth of the new states of the oscillators. Fig.\ref{fig:CL}(c) illustrates projections of the phase trajectories for oscillators of both the coherence and incoherence domains for the case of $P=3 $ and sufficiently high value of $\sigma $. It is clearly seen that the oscillations correspond to different limit cycles. This feature of the oscillators of incoherence domain is similar with the feature of solitary states, which are observed in different systems with attractive coupling \cite{jaros2018solitary,maistrenko2014solitary,teichmann2019solitary,shepelev2017solitary,rybalova2017transition,semenova2018mechanism}, including the lattice of van der Pol oscillators \cite{shepelev2020role}. Thus, this regime can be considered as a type of solitary state. When the value of $P $ is larger and the coupling strength is weaker (see fig.\ref{fig:CL}(d)), the limit cycles corresponding to oscillations in coherence and incoherence clusters becomes significantly more similar. They are depicted in fig.\ref{fig:CL}(f).

\subsection{Chimera states}

\begin{figure}[!ht]
\centering
\hspace{-1mm}\parbox[c]{.334\linewidth}{ 
\center $P=5 $
  \includegraphics[width=\linewidth]{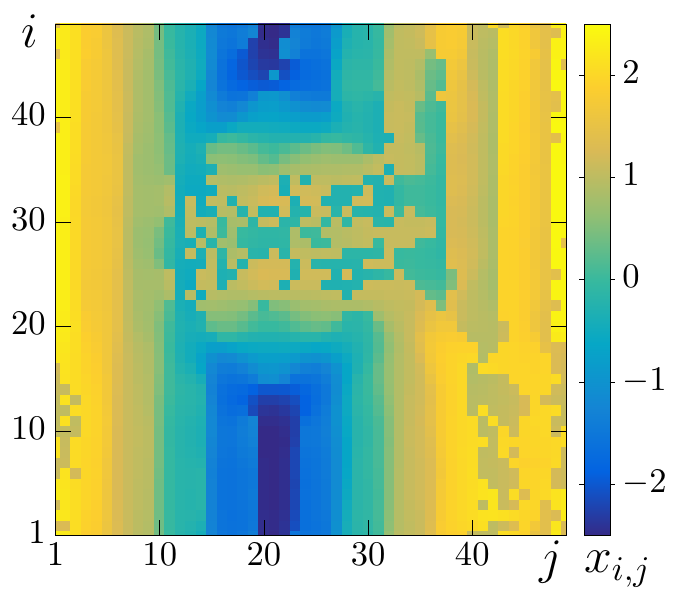}
    \vspace{-9.5mm} \center (a)
      \includegraphics[width=\linewidth]{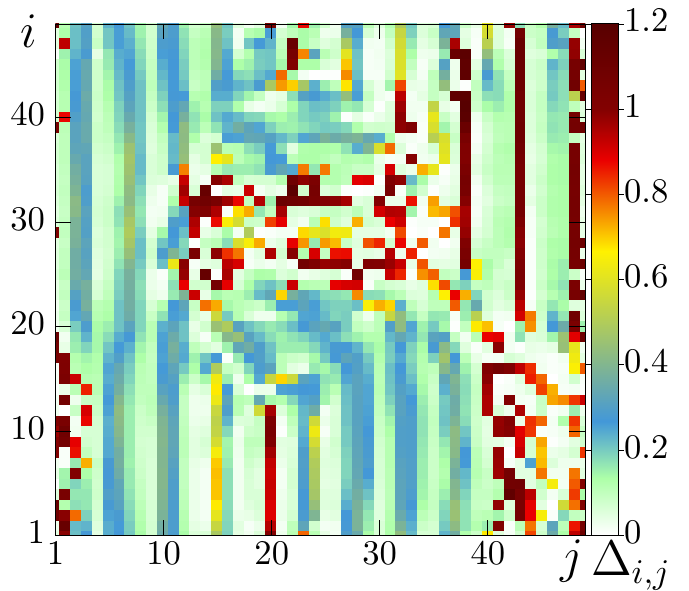}
    \vspace{-9.5mm} \center (b)
          \includegraphics[width=\linewidth]{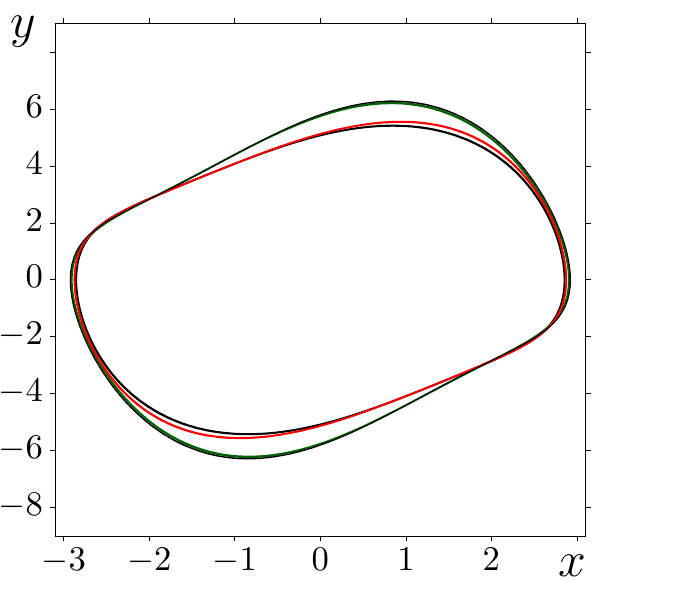}
    \vspace{-9.5mm} \center (c)
}
\hspace{-1mm}\parbox[c]{.334\linewidth}{
\center $P=6 $
\includegraphics[width=\linewidth]{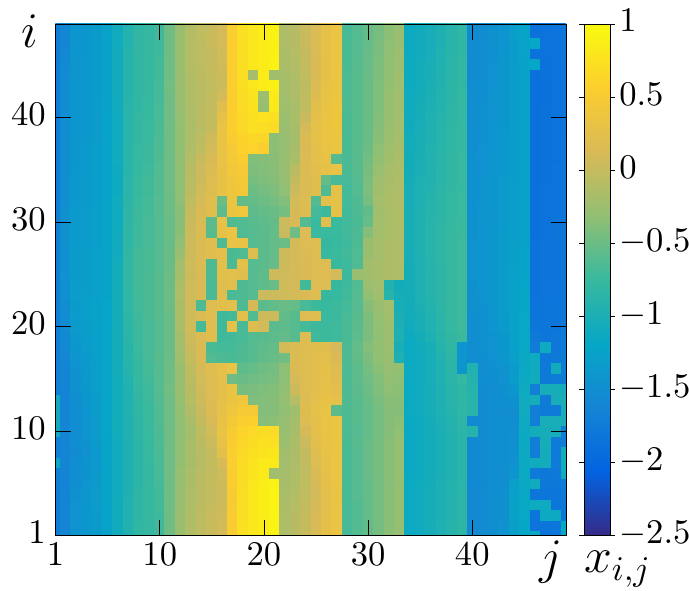}
\vspace{-9.5mm}\center (d)
  \includegraphics[width=\linewidth]{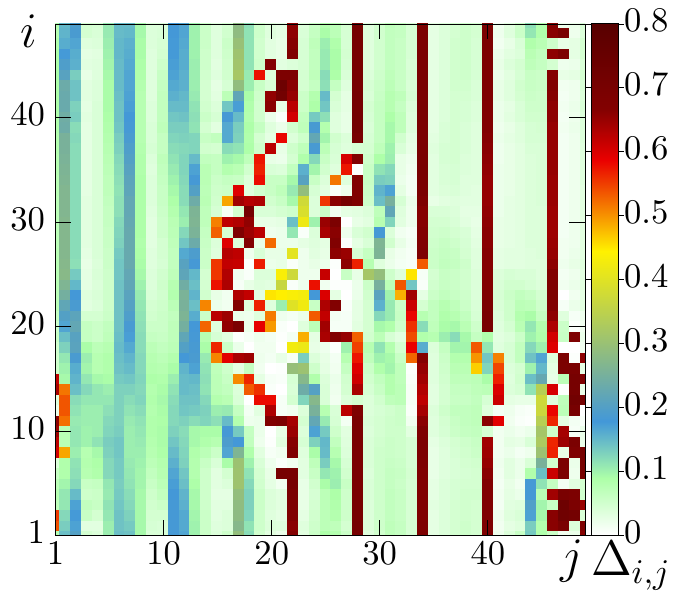}
    \vspace{-9.5mm} \center (e)
          \includegraphics[width=\linewidth]{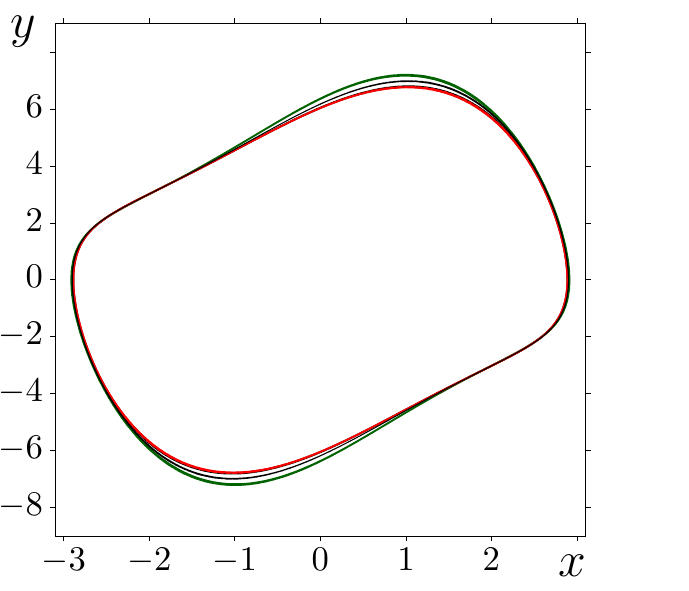}
    \vspace{-9.5mm} \center (f)
}
\hspace{-1mm}\parbox[c]{.334\linewidth}{
\center $P=8 $
\includegraphics[width=\linewidth]{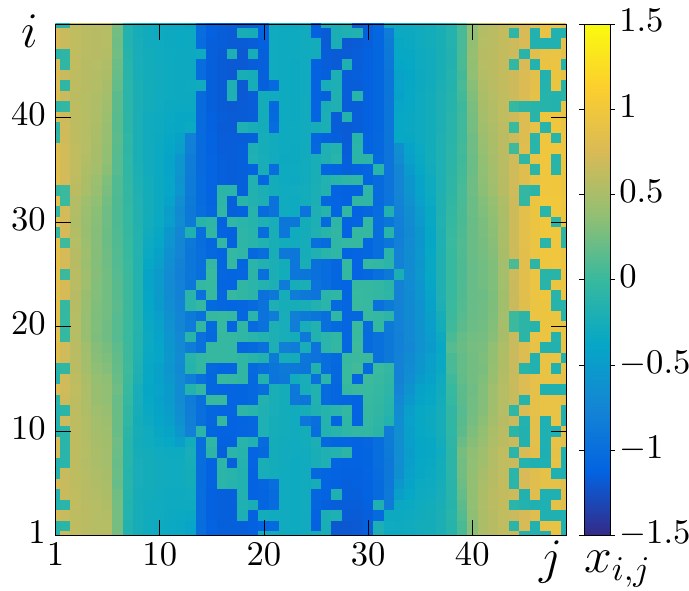}
\vspace{-9.5mm}\center (g)
  \includegraphics[width=\linewidth]{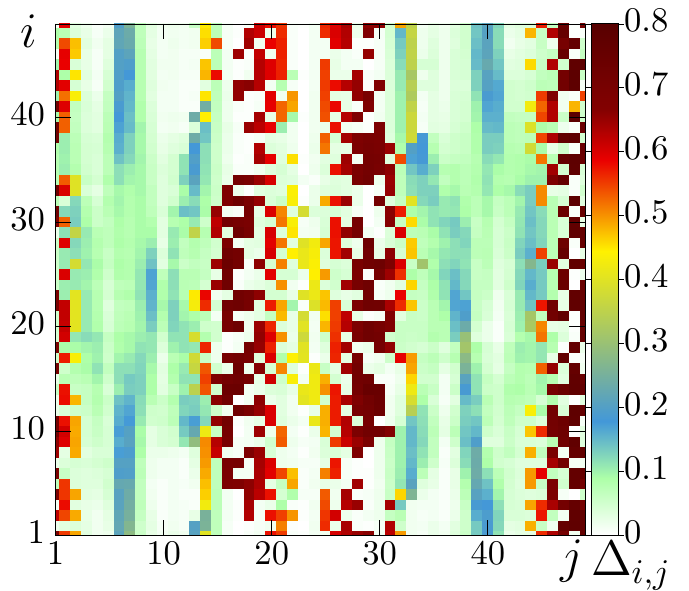}
    \vspace{-9.5mm} \center (h)
          \includegraphics[width=\linewidth]{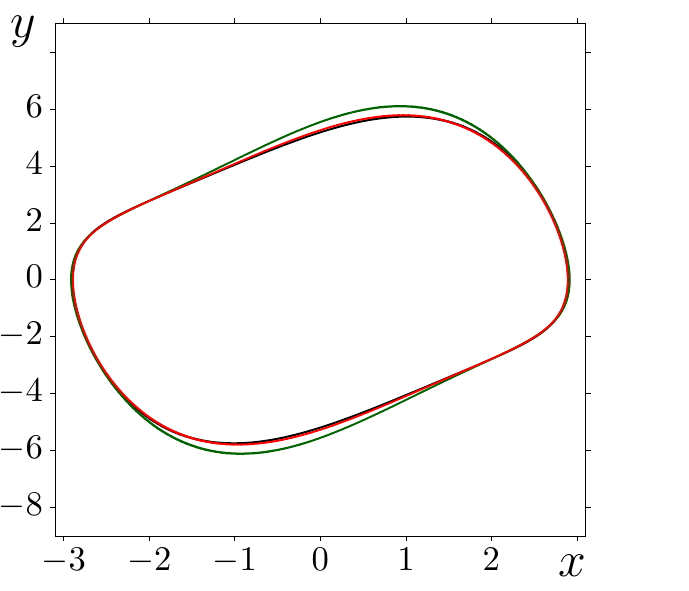}
    \vspace{-9.5mm} \center (i)
}
\caption{(Color online) Chimera states in the system \eqref{eq:grid} (region $CH $ in the regime diagram in fig.\ref{fig:regime_map}). The case of $P=5 $ and $\sigma=0.9 $ is illustrated by panels  (a)-(c), the case of $P=6 $ and $\sigma=0.5 $ corresponds to panels (d)-(e), and the case of $P=8 $ and $\sigma=0.96 $ corresponds to panels (g)-(h). (a), (d) and (g) are snapshots of the system states; (b), (e), (h) are spatial distributions of the RMSD,  projections of the phase trajectories for (c) the  $i=10, j=22 $th (solid black line), $i=31, j=26 $th (solid green line), $i=31, j=27 $th (solid red line); for (f) the  $i=26, j=18 $th (solid black line), $i=27, j=18 $th (solid green line), $i=39, j=13 $th (solid red line); for (i) the  $i=21, j=24 $th (solid black line), $i=22, j=24 $th (solid green line), $i=26, j=10 $th (solid red line). Parameters: $m=0.02 $, $\varepsilon=2.1 $, $\omega=2.5$, $N=50 $.} 
\label{fig:CH}
\end{figure}

A further increase in the coupling range leads to the growth of a number of oscillators with the asynchronous behavior. They tend to form an incoherence cluster of the large size, localized in certain places of the lattice. Thus, we can consider this state as a chimera. Examples of the chimera states in the system \eqref{eq:grid} are illustrated in figs.\ref{fig:CH}(a),(d) and (c) for different values of $P$.
All spatiotemporal structures contain two sufficiently large incoherence clusters. As it has been mentioned above, the square topology of structures is already absent for these values  of $P $ for initial conditions under study. In order to quantify this regime we calculate the spatial distribution of the root-mean-square deviation (RMSD) by the following formula:
\begin{equation}
\begin{array}{l}
\Delta_{i,j}=\sqrt{\langle(x_{i,j}-x_{i+1,j+1})^2\rangle},
\end{array}
\label{eq:RMSD}
\end{equation}
where $\langle \rangle $ mean averaging in time.

This characteristic shows statistical difference between states of adjacent oscillators. If the instantaneous amplitude and phases of oscillations of adjacent elements are similar then the RMSD values are small. At the same time, values of $\Delta_{i,j} $ are large when the oscillation features of two adjacent elements are noticeably different. Thus, the coherence cluster is characterized by small values of the RMSD, and the incoherence cluster has high values of the $ \Delta_{i,j}$. Spatial distributions of the RMSD for the structures under study are represented in figs.\ref{fig:CH}(b),(e) and (h). The incoherence clusters for all the three are characterized by the maximum values of the RMSD, while the coherence clusters have sufficiently low values of $ \Delta_{i,j}$. Besides, there are vertical lines in figs.\ref{fig:CH}(b) and (e) with high values of $ \Delta_{i,j}$. Their existence is associated with a step-like shape of the spatial structures of coherence clusters, which is seen from the snapshots of system states in figs.\ref{fig:CH}(a) and (d). The instantaneous states of adjacent oscillators in the boundaries of these "steps" are always significantly different and values of the RMSD for these oscillators are large. Interestingly, for the long coupling range ($P=8 $) the step-like shape disappears and a structure within the coherence cluster becomes smooth. 
As it has been shown above a reason of the emergence of an incoherence in the system \eqref{eq:grid} is the appearance of new stable periodic states with growing the $P $ and $\sigma $. We assume that the elongation of the coupling range leads to an expansion of the basin of attraction of the new periodic regimes.
 Hence, more and more number of oscillators are inside these basins from randomly distributed initial conditions. Coexistence of different limit cycles are illustrated by the $(x_{i,j},y_{i,j}) $ projections of the phase trajectories for different oscillators in figs.\ref{fig:CH}(c),(f) and (i). There are at least three different limit cycles of individual oscillators in the common $(x,y) $ plane for the cases $P=5 $ and $P=6 $. At the same time, difference between limit cycles when $P=6 $ is less visible than when $P=5 $. Apparently, this is due to the fact that the value of the coupling strength $\sigma $ is significantly lower for the case of $P=6 $. When the coupling range becomes longer ($P=8$), only two characterized periodic attractors remains in the system (see fig. {fig:CH}(i)). We assume that this is a consequence of the strong influence of the coupling nonlinearity. As it has been shown above, elongation of $P$ leads to a significant decrease in the linear repulsive coupling term and the reinforcement of the nonlinear attractive coupling term. Hence, when the nonlinear attractive coupling term prevails over the linear one, the system under study becomes two-state with two distinctive stable periodic attractors in the phase plane of individual oscillators. Thus, the chimera states in the lattice \eqref{eq:grid} are similar to solitary state chimeras, which exist in a lattice of van der Pol oscillators with the attractive coupling \cite{shepelev2020role} and other systems \cite{rybalova2018mechanism,rybalova2019solitary,mikhaylenko2019weak}.

\section{Two-state irregular structures}

\begin{figure}[!ht]
\centering
\hspace{-1mm}\parbox[c]{.334\linewidth}{ 
  \includegraphics[width=\linewidth]{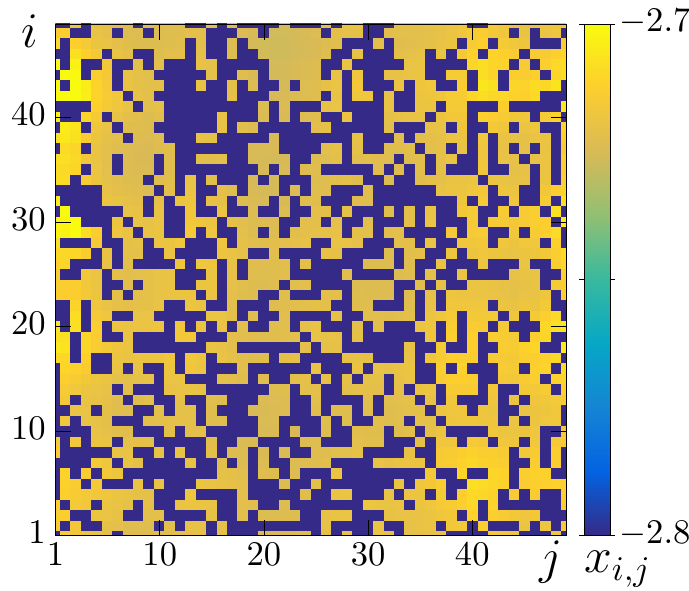}
    \vspace{-9.5mm} \center (a)
}
\hspace{-1mm}\parbox[c]{.334\linewidth}{
\includegraphics[width=\linewidth]{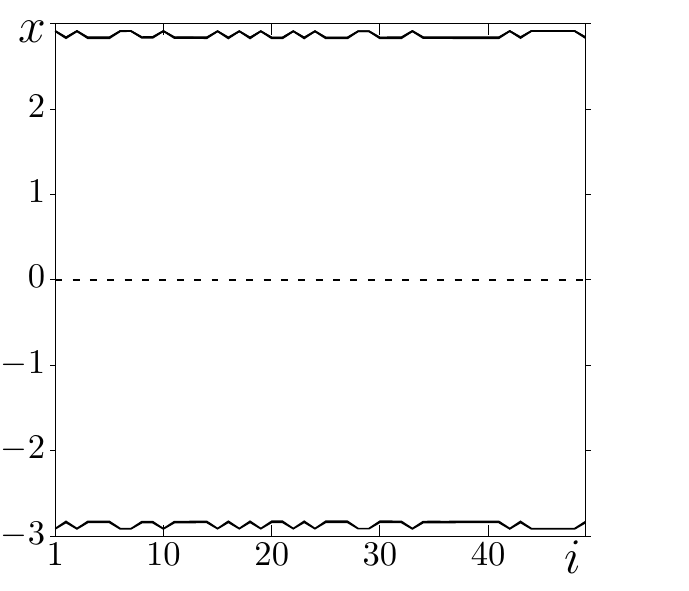}
\vspace{-9.5mm}\center (b)
}
\hspace{-1mm}\parbox[c]{.334\linewidth}{
\includegraphics[width=\linewidth]{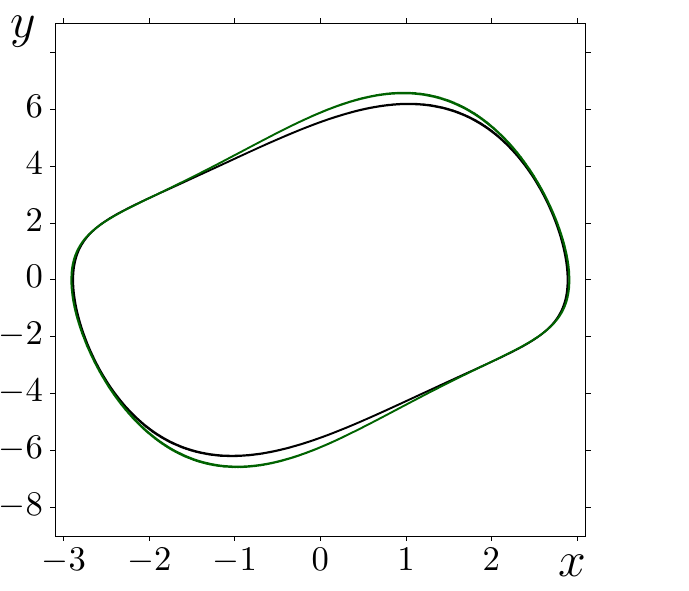}
\vspace{-9.5mm}\center (c)
}
\caption{(Color online) Two-state irregular structures (region $TS $ in the regime diagram in fig.\ref{fig:regime_map}) when $P=9 $ and $\sigma=0.74 $. (a) snapshot of the system state, (b) spatial crossection for $j=24 $ shown every half period $T/2 $, (c) phase portrait projection for the $i=22,j=25 $th (black line) and for the $i=23,j=25 $th (green line). Parameters: $m=0.02 $, $\varepsilon=2.1 $, $\omega=2.5$, $N=50 $.} 
\label{fig:BS}
\end{figure}

Now we discover the system dynamics when the coupling range is sufficiently long. As it has been mentioned in the previous section, the system under study becomes two state (there are only two different state of each oscillators) when the value of $P $ is large, namely there are only two coexisting limit cycles. A special spatiotemporal behavior is observed in region $TS $ of the regime diagram in fig.\ref{fig:regime_map}. This is the only possible regime when $P >13 $. An example of this state is depicted in fig.\ref{fig:BS}(a). The structure represents irregularly distribution between two the states with two close certain levels. The spatiotemporal dynamics is shown by a set of snapshots of the system states shown after every half period $T/2 $ in fig.\ref{fig:BS}(b). The features of this regime is similar to the case of complete synchronization. However, there are two peculiar states, and all the elements are irregularly distributed between them. These levels correspond to the two stable periodic attractors of individual oscillators, which are shown in the projections of the phase trajectories for different selected oscillators in fig.\ref{fig:BS}(c). Apparently, the increasing influence of the nonlinear attractive coupling term leads to the similarity of the sizes of the two basins  of attraction.  Hence, the oscillators are equiprobably located between them.

\section{Influence of the coupling nonlinearity parameter on the system dynamics}

\begin{figure}[!ht]
\centering
  \includegraphics[width=0.6\linewidth]{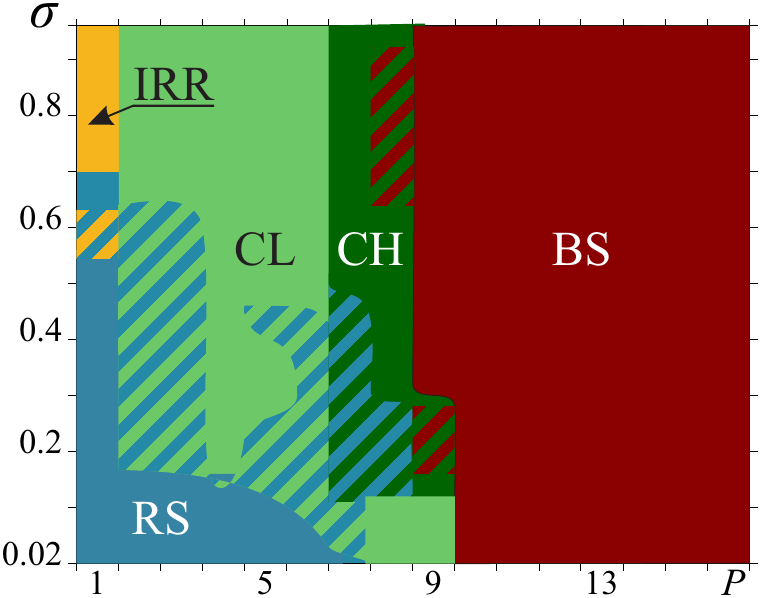}
\caption{(Color online) Diagram of the regimes for the lattice \eqref{eq:grid} in the ($P,~\sigma$) parameter plane when $m=0.06 $.
Region IRR corresponds to complete incoherence; region RS relates to the regime of regular structures; CL is the region of existence of chimera-like structures; CH correspond to chimera states; TS is the region of  two-state structures. The regions of coexistence of different regimes are shown by alternating strips of the corresponding colors (tones). Parameters: $\varepsilon=2.1 $, $\omega=2.5$, $N=50 $.}
\label{fig:regime_map_m_0_06}
\end{figure}

In the previous sections, the coefficient of nonlinear attractive coupling in \eqref{eq:grid} has been fixed as $m=0.02 $ and we have varied the parameters of the linear repulsive coupling term. Now a question arises, how does an increase in the coefficient $m$ change the system dynamics? To answer this question, we study a variety of the main dynamical regimes when a value of the parameter $m $ is fixed as $m=0.06 $. Fig.\ref{fig:regime_map_m_0_06} demonstrates a diagram of the main regimes in the ($P,~\sigma $) parameter plane.
For this value of $m $ there are the same dynamical regimes as for the previous case of $m=0.02 $. At the same time, the regions of existence of the different regimes has noticeably changed. At first, the irregular structures in the region $IRR$ are observed only for the case of local coupling ($P=1 $) and higher values of the coupling strength $\sigma $ than for the case of $m=0.02 $. The region of chimera states (region $CH $) becomes more narrow. However, the main difference is that the regime of two-state structures (region $TS $) is now realized for the significantly shorter coupling range $P $. These changes can be explained by the stronger influence of the nonlinear coupling term in the present case. Hence, all the effects associated with the coupling nonlinearity are realized for lower values of $P $ and $\sigma $. 

\begin{figure}[!ht]
\centering
  \includegraphics[width=0.6\linewidth]{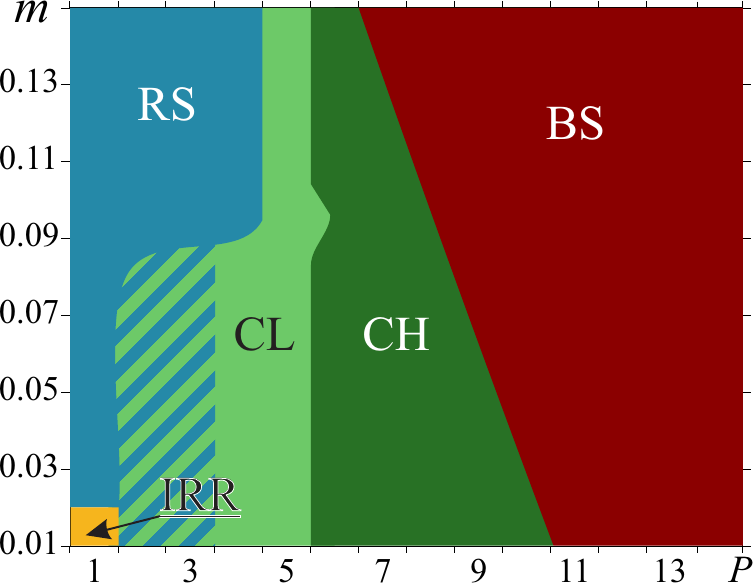}
\caption{(Color online) Phase-parametric diagram of the regimes for the lattice \eqref{eq:grid} in the ($P,~m$) parameter plane for the fixed value of $\sigma=0.3 $.
Region IRR corresponds to complete incoherence; region RS relates to the regime of regular structures; CL is the region of existence of chimera-like structures; CH correspond to chimera states; TS is the region of  two-state structures. The regions of coexistence of different regimes are shown by alternating strips of the corresponding colors (tones). Parameters: $m=0.02 $, $\varepsilon=2.1 $, $\omega=2.5$, $N=50 $.}
\label{fig:diagram_P-m}
\end{figure}

Now we fix the value of the coupling strength as $\sigma=0.3 $ and vary the coupling range $P $ and the nonlinear coupling coefficient $m $. All the main regimes are observed for the chosen value of $\sigma $. This study enables us to evaluate how the system dynamics has changed with an increase in the nonlinearity of coupling. Fig.\ref{fig:diagram_P-m} demonstrate the parametric diagram of regimes in the ($P,~m $) parameter plane.
This diagram shows that the regime of irregular structures are possible only for the minimum values of the nonlinear coupling coefficient $m $. The region of regular structures (region $RS $) is realized for the short coupling range and expands with the growth of $m $. Elongation of $P $ leads to the transition to the region of chimera-like states, which, in turn, transform into chimera structures with large incoherence clusters. Both the regions narrow down with an increase in the coupling nonlinearity, unlike region $RS $. A further increase in the values of the coupling range $P $ leads to a switch of the system \eqref{eq:grid} to the regime of two-state structures (region $TS $). Growth of the coefficient $m $ is accompanied by almost linear expansion of this region. Apparently, the increase in the coefficient of nonlinear attractive coupling $m $ decreases the irregularity in the system and restricts a number of coexisting periodic states of individual elements. For this reason, chimera and chimera-like states are realized within a wide range of $P $ values only when the coupling nonlinearity is sufficiently weak. The strong coupling nonlinearity leads to the coexistence of two limit cycles of individual elements, which are characterized by the similar sizes of the basins of attractions.

\section*{Conclusions}

In this work we have studied the dynamics of a two-dimensional lattice of nonlocally interacting identical van der Pol oscillators with linear repulsive and nonlinear attractive coupling between the elements with toroidal boundary conditions. We derive the system equation from a radiophysical circuit of the coupled van der Pol oscillators interacting through the nonlinear element with a current-voltage characteristic of the N type (for example, a tunnel diode). The behavior of the 2D lattice under study is discovered for both the local and nonlocal character of an oscillator interaction and for variation of the coupling parameters. The coupling function can be divided into two terms, namely the linear repulsive coupling and nonlinear attractive coupling. Moreover, the coefficient of nonlinear coupling does not depend on values of coupling strength $\sigma $ of linear coupling and of the coupling range $P $. This character of coupling extremely changes the dynamics of the system and the dynamical regimes becomes completely different from the case of a 2D lattice of van der Pol oscillators with common attractive (dissipative)  coupling \cite{bukh2019bspiral,shepelev2020role}. For example, all the regimes represent standing waves for any randomly distributed initial conditions under consideration. At the same time, there are no any regimes of traveling, spiral and target waves, and of spatio-uniform structures in the lattice with both the local and nonlocal  interaction.

We have constructed the  diagrams of the regimes, that are realized in the lattice \eqref{eq:grid}, in the ($P,~\sigma $) parameter plane when the nonlinear coupling coefficient is fixed as $m=0.02 $ and $m=0.06 $. Numerical simulation of the lattice dynamics enables us to reveal the following features of its behavior.
When the coupling strength is sufficiently weak and the coupling range is short, different structures with  the regular temporal behavior form in the lattice. When the coupling is local, the shape of the structures are very complex. The coupling nonlocality leads to significant change  in the structure shape, which becomes simpler and has the square step-like shape, namely groups of oscillators forms rectangles of a certain width, where all the instantaneous phases and amplitudes are very similar. At the same time oscillations of all the elements are characterized by the same periodic attractor in the phase plane of an individual oscillator. An increase in the coupling strength induces the emergence of new stable periodic states for individual oscillators in the system \eqref{eq:grid}. However, the elongation of the coupling range restricts this phenomenon due to growth of the influence of nonlinear attractive coupling. For this reason, this effect is mostly strong when the coupling range is equal $P=1 $ or $P=2 $. Emergence of a large number of coexisting stable limit cycles for different elements increases the spatiotemporal incoherence and the spatial structures become irregular. As it has been mentioned, elongation of the coupling range decreases the number of coexisting stable states. This leads to the appearance of new types of spatiotemporal structures, namely the chimera-like and chimera states. They are characterized by the coexistence of clusters of the synchronous (coherence) and asynchronous (incoherence) behavior of neighbor oscillators. The number of elements with incoherent behavior are very small for chimera-like structures and is sufficiently large for chimeras. An important feature of this structures is that oscillations of all the  coherence cluster elements corresponds to the same periodic attractor in the $(x,y) $ plane, while elements of the incoherence cluster are characterized by several other coexisting periodic attractors. Thus, these chimeras are similar to solitary state chimeras, observed in systems with the attractive coupling \cite{rybalova2018mechanism,rybalova2019solitary,mikhaylenko2019weak,shepelev2020role}.  We explore that the chimera-like states evolve into the chimeras with elongation of the coupling range. For this regime, all the oscillator are irregularly distributed between two stable states. 

Additionally, we study the behavior of the system when the nonlinear coupling coefficient  $ m$ is varied  and the coupling strength is fixed. We plot the regime diagram in the ($P,~m $) parameter plane. The diagram shows that the region of chimera and chimera-like structures decrease with the growth of $m $. On the contrary, the regions with simple spatial structures and two-state structures increase. Thus, the nonlinear attractive coupling leads to the simpler behaviour of the lattice under study.

\section*{Acknowledgements}
\vspace{0.5cm}
I.A.S. and T.E.V. thank for the financial support the Russian Science Foundation (grant 20-12-00119), S.S.M. acknowledges the use of New Zealand eScience Infrastructure (NeSI) high performance computing facilities as part of this research.

\bibliographystyle{plain} 
\bibliography{mybib}

\begin{thebibliography}{10}

\bibitem{Abrams-2004}
D.M. Abrams and S.H. Strogatz.
\newblock Chimera states for coupled oscillators.
\newblock {\em Physical review letters}, 93(17):174102, 2004.

\bibitem{Astakhov-2016}
S.~Astakhov, A.~Gulai, N.~Fujiwara, and J.~Kurths.
\newblock The role of asymmetrical and repulsive coupling in the dynamics of
  two coupled van der pol oscillators.
\newblock {\em Chaos}, 26:023102, 2016.

\bibitem{Balazsi-2001}
G.~Bal\'{a}zsi, A.~Cornell-Bell, A.B. Neiman, and F.~Moss.
\newblock Synchronization of hyperexcitable systems with phase-repulsive
  coupling.
\newblock {\em Phys. Rev. E.}, 64:041912, 2001.

\bibitem{Barrat-2008}
A.~Barrat, M.~Barthelemy, and A.~Vespignani.
\newblock {\em Dynamical processes on complex networks}.
\newblock Cambridge university press, 2008.

\bibitem{Belych-2018}
I.~Belykh, D.~Carter, and R.~Jeter.
\newblock Synchronization in multilayer networks: When good links go bad.
\newblock {\em SIAM Journal on Applied Dynamical Systems}, 18(4):2267--2302,
  2019.

\bibitem{Bera-2016}
B.K. Bera, C.~Hens, and D.~Ghosh.
\newblock Emergence of amplitude death scenario in a network of oscillators
  under repulsive delay interaction.
\newblock {\em Phys. Lett. A. 2016}, 380(31–32):2366–2373, 2016.

\bibitem{Bera-2017}
B.K. Bera, S.~Majhi, D.~Ghosh, and M.~Perc.
\newblock Chimera states: Effects of different coupling topologies.
\newblock {\em EPL}, 118:1, 2017.

\bibitem{Boccalett-2018}
S.~Boccaletti, A.~Pisarchik, C.~Genio, and A.~Amann.
\newblock {\em Synchronization: From Coupled Systems to Complex Networks}.
\newblock CUP, 03 2018.

\bibitem{bukh2019bspiral}
A.~Bukh, G.~Strelkova, and V.~Anishchenko.
\newblock Spiral wave patterns in a two-dimensional lattice of nonlocally
  coupled maps modeling neural activity.
\newblock {\em Chaos, Solitons \& Fractals}, 2019.
\newblock (in press).

\bibitem{Chen-2009}
Y.~Chen, J.~Xiao, W.~Liu, L.~Li, and Y.~Yang.
\newblock Dynamics of chaotic systems with attractive and repulsive couplings.
\newblock {\em Phys. Rev. E.}, 80:046206, 2009.

\bibitem{Genio-2016}
C.I. del Genio, J.~G{\'o}mez-Garde{\~n}es, I.~Bonamassa, and S.~Boccaletti.
\newblock Synchronization in networks with multiple interaction layers.
\newblock {\em Science Advances}, 2(11), 2016.

\bibitem{Kaue-2012}
T.~Kau\ {e}, D.~Peron, and F.A. Rodrigues.
\newblock Explosive synchronization enhanced by time-delayed coupling.
\newblock {\em Phys. Rev. E.}, 86:016102, 2012.

\bibitem{Hagerstrom-2012}
A.M. Hagerstrom, T.E. Murphy, R.~Roy, P.~H{\"o}vel, I.~Omelchenko, and
  E.~Sch{\"o}ll.
\newblock Experimental observation of chimeras in coupled-map lattices.
\newblock {\em Nature Physics}, 8:658--661, 2012.

\bibitem{Hens-2013}
C.R. Hens, O.I. Olusola, P.~Pal, and S.K. Dana.
\newblock Oscillation death in diffusively coupled oscillators by local
  repulsive link.
\newblock {\em Phys. Rev. E.}, 88(3):034902, 2013.

\bibitem{Hens-2014}
C.R. Hens, P.~Pal, S.K. Bhowmick, P.K. Roy, A.~Sen, and S.K. Dana.
\newblock Diverse routes of transition from amplitude to oscillation death in
  coupled oscillators under additional repulsive links.
\newblock {\em Phys. Rev. E.}, 89(3):032901, 2014.

\bibitem{Hong-2011}
H.~Hong and S.H. Strogatz.
\newblock Kuramoto model of coupled oscillators with positive and negative
  coupling parameters: An example of conformist and contrarian oscillators.
\newblock {\em PRL}, 106:054102, 2011.

\bibitem{jaros2018solitary}
P.~Jaros, S.~Brezetsky, R.~Levchenko, D.~Dudkowski, T.~Kapitaniak, and
  Y.~Maistrenko.
\newblock Solitary states for coupled oscillators with inertia.
\newblock {\em Chaos: An Interdisciplinary Journal of Nonlinear Science},
  28(1):011103, 2018.

\bibitem{Kuramoto-1984}
Y.~Kuramoto.
\newblock Chemical oscillations, waves and turbulence.
\newblock {\em Springer-Verlag}, 1984.

\bibitem{Kuramoto-2002}
Y~Kuramoto and D~Battogtokh.
\newblock Coexistence of coherence and incoherence in nonlocally coupled phase
  oscillators.
\newblock {\em Nonlinear Phenom. Complex Syst.}, 5(4):380--385, 2002.

\bibitem{Kuznetsov-2009}
A.P. Kuznetsov, N.V. Stankevich, and L.V. Turukina.
\newblock Duffing oscillators: phase dynamics and structure of synchronization
  tongues.
\newblock {\em Physica D}, 238:14:1203--1215, 2009.

\bibitem{Liu-2007}
X.~Liu and T.~Chen.
\newblock Exponential synchronization of nonlinear coupled dynamical networks
  with a delayed coupling.
\newblock {\em Physica A: Statistical Mechanics and its Applications},
  381:82--92, 2007.

\bibitem{Maistrenko-2014}
Y.~Maistrenko, B.~Penkovsky, and M.~Rosenblum.
\newblock Solitary state at the edge of synchrony in ensembles with attractive
  and repulsive interactions.
\newblock {\em Phys. Rev. E.}, 89:060901, 2014.

\bibitem{maistrenko2014solitary}
Y.~Maistrenko, B.~Penkovsky, and M.~Rosenblum.
\newblock Solitary state at the edge of synchrony in ensembles with attractive
  and repulsive interactions.
\newblock {\em Physical Review E}, 89(6):060901, 2014.

\bibitem{Martens-2013}
E.A. Martens, S.~Thutupalli, A.~Fourri{\`e}re, and O.~Hallatschek.
\newblock Chimera states in mechanical oscillator networks.
\newblock {\em Proceedings of the National Academy of Sciences},
  110(26):10563--10567, Jun 2013.

\bibitem{mikhaylenko2019weak}
M.~Mikhaylenko, L.~Ramlow, S.~Jalan, and A.~Zakharova.
\newblock Weak multiplexing in neural networks: Switching between chimera and
  solitary states.
\newblock {\em Chaos: An Interdisciplinary Journal of Nonlinear Science},
  29(2):023122, 2019.

\bibitem{Mishra-2015}
A.~Mishra, C.~Hens, M.~Bose, P.K. Roy, and S.K. Dana.
\newblock Chimeralike states in a network of oscillators under attractive and
  repulsive global coupling.
\newblock {\em Phys. Rev. E.}, 92(6):062920, 2015.

\bibitem{Nandan-2014}
M.~Nandan, C.R. Hens, P.~Pal, and S.K. Dana.
\newblock Transition from amplitude to oscillation death in a network of
  oscillators.
\newblock {\em Chaos}, 24:043103, 2014.

\bibitem{Nekorkin-2002}
V.I. Nekorkin and M.G. Velarde.
\newblock {\em Synergetic Phenomena in Active Lattices}.
\newblock Springer Series in Synergetics. Springer, Berlin, Heidelberg, 2002.

\bibitem{Omelchenko-2011}
I.~Omelchenko, Y.~Maistrenko, P.~H{\"o}vel, and E.~Sch{\"o}ll.
\newblock Loss of coherence in dynamical networks: spatial chaos and chimera
  states.
\newblock {\em Phys. Rev. Lett.}, 106:234102, 2011.

\bibitem{Osipov-2007}
G.V. Osipov, J.~Kurths, and C.~Zhou.
\newblock {\em Synchronization in Oscillatory Networks}.
\newblock Springer Series in Synergetics. Springer, Berlin, Heidelberg, 2007.

\bibitem{Panaggio-2015}
M.J. Panaggio and D.M. Abrams.
\newblock Chimera states: coexistence of coherence and incoherence in networks
  of coupled oscillators.
\newblock {\em Nonlinearity}, 28(3):R67--R87, Feb 2015.

\bibitem{Petereit-2017}
J.~Petereit and A.~Pikovsky.
\newblock Chaos synchronization by nonlinear coupling.
\newblock {\em CNSNS}, 44:344--351, 2017.

\bibitem{Rabinovich-2006}
M.I. Rabinovich, P.~Varona, A.~I. Selverston, and H.~D.~I. Abarbanel.
\newblock Dynamical principles in neuroscience.
\newblock {\em Rev. Mod. Phys.}, 78:1213, 2006.

\bibitem{rybalova2019solitary}
E.~Rybalova, V.S. Anishchenko, G.I. Strelkova, and A.~Zakharova.
\newblock Solitary states and solitary state chimera in neural networks.
\newblock {\em Chaos: An Interdisciplinary Journal of Nonlinear Science},
  29(7):071106, 2019.

\bibitem{rybalova2017transition}
E.~Rybalova, N.~Semenova, G.~Strelkova, and V.~Anishchenko.
\newblock Transition from complete synchronization to spatio-temporal chaos in
  coupled chaotic systems with nonhyperbolic and hyperbolic attractors.
\newblock {\em Eur. Phys. J. Spec. Top.}, 226:1857--1866, Jun 2017.

\bibitem{Rybalova-2019}
E.V. Rybalova, D.Y. Klyushina, V.S. Anishchenko, and G.I. Strelkova.
\newblock Impact of noise on the amplitude chimera lifetime in an ensemble of
  nonlocally coupled chaotic maps.
\newblock {\em Regular and Chaotic Dynamics}, 24(4):432--445, 2019.

\bibitem{rybalova2018mechanism}
E.V. Rybalova, G.I. Strelkova, and V.S. Anishchenko.
\newblock Mechanism of realizing a solitary state chimera in a ring of
  nonlocally coupled chaotic maps.
\newblock {\em Chaos, Solitons \& Fractals}, 115:300--305, 2018.

\bibitem{Dixit-2019}
M.~Dev~Shrimali S.~Dixit, A.~Sharma.
\newblock The dynamics of two coupled van der pol oscillators with attractive
  and repulsive coupling.
\newblock {\em Physics Letters A.}, 383:125930, 2019.

\bibitem{Schmidt-2012}
G.S. Schmidt, A.~Papachristodoulou, U.~M\"{u}nz, and F.~Allg\"{u}wer.
\newblock Frequency synchronization and phase agreement in kuramoto oscillator
  networks with delays.
\newblock {\em Automatica}, 48:3008–3017, 2012.

\bibitem{semenova2018mechanism}
N.~Semenova, T.~Vadivasova, and V.S. Anishchenko.
\newblock Mechanism of solitary state appearance in an ensemble of nonlocally
  coupled lozi maps.
\newblock {\em The European Physical Journal Special Topics},
  227(10-11):1173--1183, 2018.

\bibitem{shepelev2020role}
I.A. Shepelev, A.V. Bukh, S.S. Muni, and V.S. Anishchenko.
\newblock Role of solitary states in forming spatiotemporal patterns in a 2d
  lattice of van der pol oscillators.
\newblock {\em Chaos, Solitons \& Fractals}, 135:109725, 2020.

\bibitem{shepelev2017solitary}
I.A. Shepelev and T.E. Vadivasova.
\newblock Solitary states in a 2d lattice of bistable elements with global and
  close to global interaction.
\newblock {\em Nelineinaya Dinamika [Russian Journal of Nonlinear Dynamics]},
  13(3):317--329, 2017.

\bibitem{Shepelev-2019}
I.A. Shepelev and T.E. Vadivasova.
\newblock Variety of spatio-temporal regimes in a 2d lattice of coupled
  bistable fitzhugh-nagumo oscillators. formation mechanisms of spiral and
  double-well chimeras.
\newblock {\em Communications in Nonlinear Science and Numerical Simulation},
  79:104925, 2019.

\bibitem{teichmann2019solitary}
E.~Teichmann and M.~Rosenblum.
\newblock Solitary states and partial synchrony in oscillatory ensembles with
  attractive and repulsive interactions.
\newblock {\em Chaos: An Interdisciplinary Journal of Nonlinear Science},
  29(9):093124, 2019.

\bibitem{Tsimring-2005}
L.S. Tsimring, N.~F. Rulkov, M.~L. Larsen, and M.~Gabbay.
\newblock Repulsive synchronization in an array of phase oscillators.
\newblock {\em Phys. Rev. Lett.}, 95:014101, 2005.

\bibitem{Ullner-2007}
E.~Ullner, A.~Zaikin, E.I. Volkov, and J.G. Ojalvo.
\newblock Multistability and clustering in a population of synthetic genetic
  oscillators via phase-repulsive cell-to-cell communication.
\newblock {\em Phys. Rev. Lett.}, 99:148103, 2007.

\bibitem{Volos-2015}
M.A. Volosyuk, A.V. Volosyuk, and N.Y. Rokhmanov.
\newblock The role of interstitial (crowdion) mass-transfer for crack
  high-temperature healing under uniaxial loading.
\newblock {\em Functional materials}, 2015.

\bibitem{Wang-2001}
Q.~Wang, G.~Chen, and M.~Perc.
\newblock Synchronous bursts on scale-free neuronal networks with attractive
  and repulsive coupling.
\newblock {\em PLoS ONE}, 6(1), 2001.

\bibitem{Xu-2018}
F.~Xu, J.~Zhang, M.~Jin, Sh. Huang, and T.~Fang.
\newblock Chimera states and synchronization behavior in multilayer memristive
  neural networks.
\newblock {\em Nonlinear Dynamics}, 94(2):775–783, 2018.

\bibitem{Yamamoto-2018}
H.~Yamamoto, S.~Kubota, F.A. Shimizu, A.~Hirano-Iwata, and M.~Niwano.
\newblock Effective subnetwork topology for synchronizing interconnected
  networks of coupled phase oscillators.
\newblock {\em Frontiers in Computational Neuroscience}, 12, 2018.

\bibitem{Yanagita-2005}
T.~Yanagita, T.~Ichinomiya, and Y.~Oyama.
\newblock Pair of excitable fitzhugh-nagumo elements: Synchronization,
  multistability.
\newblock {\em Phys. Rev. E.}, 72:056218, 2005.

\bibitem{Yeung-1999}
M.K.S. Yeung and S.H. Strogatz.
\newblock Time delay in the kuramoto model of coupled oscillators.
\newblock {\em Phys. Rev.Lett.}, 82:648–651, 1999.

\bibitem{Yang-2016}
Y.~Yuan, T.~Solis-Escalante, M.~van~de Ruit, F.C.T. van~der Helm, and S.C.
  Alfred.
\newblock Nonlinear coupling between cortical oscillations and muscle activity
  during isotonic wrist flexion.
\newblock {\em Frontiers in Computational Neuroscience}, 10, 2016.

\bibitem{Zakharova-2020}
A.~Zakharova.
\newblock {\em Chimera Patterns in Networks: Interplay between Dynamics,
  Structure, Noise, and Delay}.
\newblock Springer International Publishing, 2020.

\bibitem{Zakharova-2014}
A.~Zakharova, M.~Kapeller, and E.~Sch{\"o}ll.
\newblock Chimera death: Symmetry breaking in dynamical networks.
\newblock {\em Physical review letters}, 112(15):154101, 2014.

\bibitem{Zhao-2018}
N.~Zhao, Z.~Suna, and W.~Xu.
\newblock Amplitude death induced by mixed attractive and repulsive coupling in
  the relay system.
\newblock {\em Eur. Phys. J. B.}, 91:20, 2018.

\end{thebibliography}
\end{document}